\documentclass[
    oneside,
    aps,
    pra,
    superscriptaddress,
    twocolumn,
    a4paper,
    floatfix,
    longbibliography,
    10pt
]{revtex4-2}

\usepackage[american]{babel}
\usepackage[utf8x]{inputenc}
\usepackage[T1]{fontenc}

\usepackage{grffile}

\usepackage{newtxtext,newtxmath}
\usepackage{microtype}

\usepackage{amsmath}
\usepackage{mathrsfs}

\usepackage{bbm}
\usepackage{bm}
\usepackage{mathtools}
\usepackage{dsfont}
\usepackage{braket}
\usepackage{cancel}
\usepackage{slashed}
\usepackage{upgreek}

\usepackage{graphicx}
\usepackage[svgnames]{xcolor}

\usepackage{siunitx}
\DeclareSIUnit\Gal{Gal}

\usepackage{makecell}
\usepackage{booktabs}

\usepackage{ragged2e}

\usepackage{ragged2e}
\usepackage{array}
\usepackage{tabularx}
\usepackage{booktabs}

\definecolor{cset-aps-blueberry}{RGB}{28,128,158}
\definecolor{cset-aps-blue}{RGB}{46,44,184}
\definecolor{cset-aps-turquoise}{RGB}{0,67,88}
\definecolor{cset-aps-limegreen}{RGB}{190,219,67}
\definecolor{cset-aps-green}{RGB}{31,138,112}
\definecolor{cset-aps-yellow}{RGB}{255,225,25}
\definecolor{cset-aps-orange}{RGB}{253,116,0}
\definecolor{cset-aps-red}{RGB}{219,0,43}

\definecolor{blau}{HTML}{1575B9}
\definecolor{hellblau}{HTML}{65B7EF}
\definecolor{rot}{HTML}{E31B0A}
\definecolor{hellrot}{HTML}{FC6761}
\definecolor{gruen}{HTML}{25A131}
\definecolor{lila}{HTML}{2E2CB8}
\definecolor{grau}{HTML}{A6A6A6}

\usepackage{pgfplots}
\usepackage{tikz}

\pgfplotsset{compat=1.18}
\usetikzlibrary{
    shapes.geometric,
    arrows.meta,
    positioning,
    quotes,
    patterns,
    decorations.pathmorphing,
    shadows,
    fit,
    calc
}
\pgfplotsset{
    every axis legend/.append style={
        cells={anchor=west},
        at={(0.96,0.04)},
        anchor=south east,
        font=\scriptsize,
        },
    every axis/.append style={
        },
    xmajorgrids=true,
    xminorgrids=false,
    minor x tick num=1,
}
\usetikzlibrary{decorations}
\usetikzlibrary{calc}

\usepackage{pict2e,picture}
\usepackage{diagbox}

\makeatletter
\DeclareRobustCommand{\Arrow}[1][]{
    \check@mathfonts
    \if\relax\detokenize{#1}\relax
        \settowidth{\dimen@}{$\m@th\rightarrow$}
    \else
        \setlength{\dimen@}{#1}
    \fi
    \sbox\z@{\usefont{U}{lasy}{m}{n}\symbol{41}}
    \begin{picture}(\dimen@,\ht\z@)
        \roundcap
        \put(\dimexpr\dimen@-.7\wd\z@,0){\usebox\z@}
        \put(0,\fontdimen22\textfont2){\line(1,0){\dimen@}}
    \end{picture}
}
\makeatother

\usepackage{hyperref}
\hypersetup{
    colorlinks=true,
    linkcolor={cset-aps-red},
    linkbordercolor={cset-aps-red},
    filecolor={cset-aps-orange},
    filebordercolor={cset-aps-orange},
    citecolor={cset-aps-blue},
    citebordercolor={cset-aps-blue},
    urlcolor={cset-aps-green},
    urlbordercolor={cset-aps-green},
    menucolor={cset-aps-limegreen},
    menubordercolor={cset-aps-limegreen},
    breaklinks=true,
    pdfborderstyle={/S/U/W 2},
    pdfpagemode=UseOutlines,
    pdfstartpage={1},
}

\DeclareMathOperator{\sinc}{sinc}
\renewcommand{\Im}{\operatorname{Im}}
\renewcommand{\Re}{\operatorname{Re}}

\newcommand{\ee}{\text{e}}
\newcommand{\ii}{\text{i}}
\newcommand{\dd}{\text{d}}

\newcommand{\eg}{e.\,g.,}
\newcommand{\ie}{i.\,e.,}
\newcommand{\cf}{cf.}
\newcommand{\wrt}{w.\,r.\,t.}

\newcommand{\id}{\hat{\mathds{1}}}
\newcommand{\rb}{$^{87}$Rb}     

\newcommand{\mt}{\mathrm}

\usepackage{wasysym}

\usepackage[clock]{ifsym}
\usepackage{lipsum}

\usepackage{layouts}

\newcommand{\orcid}[1]{\href{https://orcid.org/#1}{\includegraphics[width=7pt]{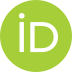}}}

\newcommand{\affTUDa}{
    \href{https://ror.org/05n911h24}{Technische Universit{\"a}t Darmstadt},
    Fachbereich Physik,
    Institut f{\"u}r Angewandte Physik,
    Schlo{\ss}gartenstr. 7,
    64289 Darmstadt,
    Germany}

\begin{document}

\title{
Balancing Quasi-Bragg Regime and Velocity Selectivity\\
in Quantum-Enhanced Atom Interferometry
}

\author{Christian M. Karres\,\orcid{0009-0005-0622-8174}\,}
\email{christian-karres@gmx.de}
\email{karresch@uni-mainz.de}
\altaffiliation{Present address: \href{https://ror.org/023b0x485}{Johannes Gutenberg Universität Mainz}, QUANTUM, Institut für Physik, Staudingerweg 7, 55128 Mainz, Germany.}
\affiliation{\affTUDa}

\author{Daniel Derr\,\orcid{0000-0002-8690-3897}\,}
\affiliation{\affTUDa}
\author{Enno Giese\,\orcid{0000-0002-1126-6352}\,}
\affiliation{\affTUDa}

\begin{abstract}
Spin squeezing in atomic ensembles enables atom interferometry with sensitivities  below the shot-noise limit, but the associated entanglement is highly susceptible to loss, making imperfections in atom optics a central limitation.
Bragg diffraction is an established technique for driving transitions between atomic momentum states and enables large-momentum transfer through higher-order diffraction while preserving the internal state.
However, it is intrinsically limited by two competing mechanisms:
Short light pulses induce parasitic diffraction into off-resonant orders beyond an effective two-level description, while long pulses face velocity selectivity. 
We derive analytical expressions in a second-quantized framework for the atom optics and phase uncertainty of a Mach--Zehnder interferometer including these effects.
We demonstrate that sub-shot-noise scaling is achieved only in a regime of intermediate pulse duration.
Furthermore, we show that deleterious effects of higher-order diffraction are partially mitigated by optimizing the input quantum state.

\vspace{2mm}\noindent This article has been published in \href{https://doi.org/10.1103/wr69-9hr5}{Physical Review Research \textbf{8}, 033247 (2026)} under the terms of the \href{https://creativecommons.org/licenses/by/4.0/}{Creative Commons Attribution License 4.0 [CC BY]}.
DOI: \href{https://doi.org/10.1103/wr69-9hr5}{10.1103/wr69-9hr5}
\end{abstract}

\maketitle

\section{Introduction}
\label{sec:Intro}
Conventional strategies to enhance the sensitivity of atom interferometers for high-precision measurements rely on boosting the atom flux~\cite{Rudolph2015-Flux, Herbst2024-Flux, Savoie2028-Flux} as well as on increasing the scaling factor through longer interrogation times (\eg{} in large-baseline fountains~\cite{Schilling2020-VLBAI, Zhou2011, Kovachy2015} or microgravity environments~\cite{Muentinga2013-Bragg_AI, Lachmann2021, Elliott2023, Williams2024-ISS_AI,Barrett2016}) and higher momentum transfer~\cite{MCGuirk2000-seq_LMT, Mueller2008, Chiow2011-Bragg_seq_LMT, Rudolph2020-LMT_single_photon_trans, Gebbe2021-BO_LMT}.
Despite such improvements, the sensitivity remains bounded by the standard quantum or shot-noise limit (SNL).
To surpass this limitation, one must go beyond classical strategies and exploit the quantum nature of atoms: Nonclassical correlations in entangled states enable quantum metrology protocols that achieve sensitivities below SNL~\cite{Pezze2018-Qrev, Szigeti2021-Q_enh_AI}.

Recent advances in atom optics provide diverse schemes for large-momentum transfer. 
For example, accelerated optical lattices are routinely used to drive Bloch oscillations~\cite{Fitzek2024-BO_LMT, Gebbe2021-BO_LMT,Parker2018-FineStr_const, Kovachy2010, Clade2015, Morel2020}, while sequential schemes rely on a sequence of light pulses, each inducing single-photon~\cite{Rudolph2020-LMT_single_photon_trans}, Raman~\cite{MCGuirk2000-seq_LMT}, or Bragg~\cite{Chiow2011-Bragg_seq_LMT, Mueller2008} transitions, enabling the cumulative transfer of multiple photon momenta.
In addition to sequential pulses, Bragg diffraction, where two counterpropagating lasers couple atoms in different momentum states without altering their internal state~\cite{Martin1988-Bragg_diff, Mazzoni2015-Bragg_AI, Giese2013-Double_Bragg}, provides a direct route to higher-order resonant processes that transfer multiple photon momenta in a single interaction, which can be readily achieved by adjusting the frequency difference of the lasers. 
Higher-order transitions rely on the fact that the atomic dynamics is not confined to two momenta.
This added flexibility, however, comes at the cost of inherent imperfections:
In the \textit{quasi-Bragg regime}, short and intense light pulses (with strong coupling and high Rabi frequency) can populate off-resonant diffraction orders~\cite{Mueller2008-adiab_elim_jBragg,Siemss2020-JulianBackgrouondPhysRevA.102.033709}. 
The behavior can be understood heuristically in terms of energy-time uncertainty, as shorter pulses result in a broader resonance and allow for the population of off-resonant momenta. 
As a consequence, parasitic diffraction introduces spurious paths to an atom interferometer, degrading the interference signal~\cite{Jenewein2022-Spurious_paths_Bragg_AI}.
Considerable effort has been devoted to mitigate or suppress spurious paths, \eg{} by utilizing dichroic mirror pulses~\cite{Pfeiffer2025-DominikDchrMirr, Siemss2020-JulianBackgrouondPhysRevA.102.033709} or optimal control techniques~\cite{Wang2024, Fang2018-OTC, Wilkason2022, Rodzinka2024, Saywell2023}.

Parasitic diffraction can be suppressed by long light pulses with weak coupling and small Rabi frequency. 
In this limit, however, another detrimental effect arises:
Due to the finite momentum width of the atomic ensemble, a Doppler detuning across the whole momentum distribution causes velocity selectivity~\cite{Szigeti2012-Velocity_selectivity}. 
As a result, the light pulses do not address the entire atomic cloud, such that only a fraction of the atoms are diffracted.
These two regimes dominate at opposing limits of the coupling strength, such that optimal performance of an interferometer is achieved by balancing the competing effects of parasitic higher-order diffraction and velocity selectivity. 
Since both mechanisms act as loss channels within the interferometric sequence, their quantitative understanding is crucial.

Accounting for loss channels becomes particularly important when the interferometer employs entangled atomic states, as any imperfection can strongly impact the achievable quantum enhancement~\cite{Günther2024-squeezingenhancementlossymultipath,Goldsmith2025-Derckeff}.
Such nonclassical, entangled states can, in principle, enable sensitivities below SNL, with the phase uncertainty scaling more favorably with the total number of atoms rather than with its square root.
In cold atomic clouds, entanglement is typically created via nonlinear interactions~\cite{Pezze2018-Qrev}. 
In two-component systems this includes spin-squeezing processes, such as one-axis twisting or two-axis counter twisting~\cite{Kitagawa1993-SSS, Gross2010-first_OAT, Riedel2010-first_OAT, Szigeti2020-OAT_MZI}.
Alternatively, spin-changing collisions in spin-1 systems enable the generation of highly entangled twin-Fock states~\cite{Zhang2013-Entanglement, Lücke2011-Q_enh_Twin_Fock, Lange2018-Q-Enh}.
Beyond direct collisions, spin squeezing via light-matter coupling is achieved via cavity-mediated interactions~\cite{Leroux2010, Shankar2019, Wilson2024, Luo2025, Srivastava2026}, as well as by laser-probed quantum non-demolition measurements~\cite{Greve2022-CavitygreveEntanglement, Malia2022}.

Typically, entanglement is created between different internal states of the atoms.
However, for atom interferometers used as inertial sensors, it is necessary to coherently transfer the entanglement to distinct momentum states~\cite{Salvi2018}. 
Such transfer has been successfully demonstrated using twin-Fock states generated by spin-changing collisions~\cite{Anders2021-mom_entagnlement} and entanglement generated by two-mode squeezing has even been employed in a quantum-enhanced atomic gravimeter~\cite{Cassens2025-QEnhanced_Gravimeter}. 
Alternative approaches propose utilizing one-axis twisted (OAT) states, generated by strong interactions~\cite{Szigeti2020-OAT_MZI}, for example facilitated by delta-kick squeezing~\cite{Corgier2021}.
Current efforts even aim to integrate spin squeezing and large-momentum transfer techniques in microgravity environments to advance the development of space-born atom interferometers~\cite{INTENTASanton2024}.

However, while entanglement in atomic clouds has been successfully demonstrated, mitigating decoherence effects remains a critical challenge for the practical implementation of quantum-enhanced interferometers for inertial sensing applications. 
Velocity selectivity and parasitic diffraction associated with Bragg pulses cause rapid degradation of the signal contrast and quantum enhancement. 
Thoroughly quantifying these loss channels is essential for analyzing their effect on the interferometer sensitivity.
For instance, in interferometers driven by $n$th-order Bragg diffraction and operated with OAT states, the atom-light interaction can be effectively described using a scattering-matrix framework~\cite{Günther2024-squeezingenhancementlossymultipath} to infer expressions for the phase sensitivity.
In this approach, phase shifts and losses due to parasitic diffraction are incorporated via Landau--Zener theory~\cite{Siemss2020-JulianBackgrouondPhysRevA.102.033709}, while velocity selectivity is treated to first order in the Doppler detuning.

Here, we present a case study of a quantum-enhanced, first-order Bragg Mach--Zehnder interferometer (MZI), examining how velocity selectivity and parasitic diffraction compete to limit quantum enhancement.
In Sec.~\ref{sec:Q-Enhancement}, we introduce an analytical, perturbative~\cite{SIso2018} second-quantized description of first-order Bragg diffraction with box pulses.
Our formalism includes atom loss due to velocity selectivity as well as parasitic diffraction and enables straightforward calculation of the MZI's interference signal and phase uncertainty for entangled input states based on Gaussian uncertainty propagation. 
In particular, we fully account for the effect of velocity selectivity in our description.
In Sec.~\ref{sec:rm_background}, we introduce a detection scheme enhancing sensitivity in the velocity-selective regime by spatially cropping the detection area.
We then derive a general analytical expression for the MZI's phase uncertainty in Sec.~\ref{sec:PS_including_VS_and_PD} and analyze the performance of OAT states in Sec.~\ref{sec:Q-Enhancement_sus_to_loss}.
Our analysis demonstrates that sub-shot-noise sensitivity is achievable by balancing the velocity-selective and quasi-Bragg regimes.
We compare these perturbative findings to a numerical model.
Furthermore, we optimize the input state to identify the optimal OAT parameters, enabling enhanced sensitivity over a wider range of operating conditions, particularly in the quasi-Bragg regime.
The employed perturbation theory is detailed in Appendix~\ref{app:Pert-Bragg}.
The MZI phase uncertainty for OAT states is listed in Appendix~\ref{app:phase_uncertainty_formulas}.

\section{Second-quantized Description of a lossy Mach--Zehnder interferometer}
\label{sec:Q-Enhancement}

\subsection{Atom-Light Interaction}
\label{sec:AL-interactin-2ndQ}
\begin{figure}[htb]
    \centering
    \includegraphics{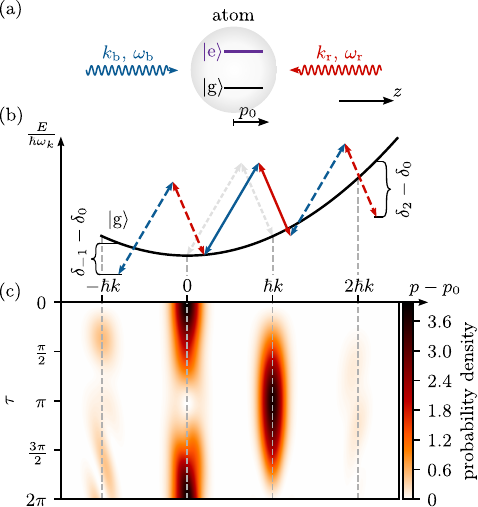}
    \caption{
        First-order Bragg diffraction.
        (a) Schematic setup for Bragg diffraction. 
        Counterpropagating lasers with wave numbers and frequencies $k_\mt{b,r}$ and $\omega_\mt{b,r}$ interrogate a two-level atom with internal energies $\hbar\omega_\mt{g}$ and $\hbar\omega_\mt{e}$. 
        (b) Kinetic energy of ground-state atoms dependent on the off-resonant momentum $p - p_0$. 
        Arrows show two-photon processes induced by the counterpropagating lasers imparting momentum $\hbar k= \hbar k_\mt{b} + \hbar k_\mt{r}$ to the atom. 
        The solid arrows depict a Doppler-detuned transition between the main momentum classes $n=0,1$. 
        The resonant coupling is shown in dashed gray, and the colored dashed arrows display the coupling to adjacent classes $n=-1,2$, which are detuned by $\delta_{-1,2} - \delta_n$ for atoms starting in class $n$.
        (c) The numerical solution of Eq.~\eqref{eq:Heisenberg_eom}, valid for all $\varepsilon$ values, gives momentum-dependent Rabi oscillations, shown here for atoms initially Gaussian distributed in momentum with mean $p_0$ and standard deviation $\sigma_q = 0.1\hbar k$.
        The Rabi frequency is set to $\Omega_0 = 0.8\omega_k$, well within the quasi-Bragg regime. 
        This choice corresponds to $\varepsilon=0.8$ and is intentionally close to the boundary of the perturbative regime to make the momentum-dependent deleterious effects clearly visible.
        Diffraction into the adjacent classes $n=-1,2$ and velocity selectivity are clearly observed.
        The latter is particularly visible at the mirror pulse with pulse area $\tau=\pi$.
        Here, only atoms with near-resonant momenta are diffracted into class $n=1$, while off-resonant atoms remain in $n=0$.
    }
    \label{fig:parabola}
\end{figure}

We consider two-level atoms of mass $m$ with energies $\hbar \omega_\mt{g}$ and $\hbar \omega_\mt{e}$, interrogated by two counterpropagating lasers as shown in Fig.~\ref{fig:parabola}(a).
The lasers have frequencies $\omega_\mt{b}$ and $\omega_\mt{r}$, and corresponding wave numbers $k_\mt{b}$ and $k_\mt{r}$.
This configuration induces two-photon transitions in which an atom absorbs a photon from one light field and subsequently undergoes stimulated emission into the counterpropagating light field.
This process transfers a total momentum of $\hbar k $ to the atom, where the effective wave number is given by $k = k_\mt{b} + k_\mt{r}$.
Since the lasers are far detuned from the atomic transition, the excited state $\ket{\mt{e}}$ is only virtually populated and the two-photon transitions couple ground-state atoms in different momentum states.
 
In the generalized process of $n$th-order Bragg diffraction, the atom undergoes $n$ consecutive two-photon transitions.
For this process to resonantly impart the momentum $n\hbar k$ to atoms with initial resonant momentum $p_0$, the detuning of the lasers $\Delta\omega = \omega_\mt{b}-\omega_\mt{r}$ has to be set to $\Delta\omega = \nu_k(p_0)+ n \omega_k$ following energy and momentum conservation, where the Doppler detuning is $\nu_k(p) = pk/m$ and the recoil frequency is $\omega_k = \hbar k^2/(2m)$.
For a convenient description, we define momentum bins $q + p_0 + n\hbar k$, with $q\in[-\hbar k/2, \hbar k/2]$ centered around $p_0 + n\hbar k$, as \textit{momentum classes}.
In this convention, $n$th-order Bragg diffraction mainly couples the zeroth to the $n$th momentum class.

To describe the momentum-dependent atom-light interaction in second quantization, we define the momentum-class field operators $\hat{\Psi}_n(q)$.
These operators describe atoms in their ground state with total initial momentum $p = p_0 + q + n\hbar k$, where $q\in[-\hbar k/2, \hbar k/2]$, \ie{} atoms in the $n$th momentum class. 
The operators satisfy the usual bosonic commutation relations
\begin{equation}
	\left[\hat{\Psi}_m^{\phantom{\dag}}(q),\hat{\Psi}_n^\dag(q')\right] = \delta_{mn}\delta(q-q')\id,\quad\
	\left[\hat{\Psi}_m(q),\hat{\Psi}_n(q')\right] = \hat{0},
\end{equation}
where $\delta_{ij}$ is the Kronecker delta and $\delta(q-q')$ is the delta distribution.
For proof of concept, we restrict our analysis to first-order Bragg diffraction, \ie{} we set $\Delta\omega = \nu_k(p_0) + \omega_k$.
This process is schematically depicted in Fig.~\ref{fig:parabola}(b) as solid arrows in the energy-momentum diagram, together with off-resonant, parasitic diffraction to adjacent momentum classes $n=-1,2$ (dashed arrows). 
To describe the atom-light interaction, we transform into a suitable interaction picture, where we identify different energy scales of processes involving the excited state and those confined to the atomic ground state.
The strong detuning of the lasers from the atomic resonance causes a separation of energy scales.
We subsequently adiabatically eliminate the coupling to the excited state, which yields effective dynamics between ground-state atoms in different momentum classes to first order in the inverse laser detuning to the atomic resonance~\cite{Sanz2016, Bott2023, Giese2015_Mechanisms, Hartmann2025}.
In this case, the system is described by the following effective Hamiltonian for resonant first-order Bragg diffraction
\begin{subequations}
\label{eq:Bragg_final_Hamiltonian}
\begin{equation}
    \hat{H} = \hbar \omega_k\sum_{m,n}\int_{I_k}\!\dd q\, \hat{\Psi}_m^\dag \mathcal{H}_{mn} \hat{\Psi}_n^{\phantom{\dag}},
\end{equation}
with
\begin{equation}
    \mathcal{H}_{nn} =  \delta_n \text{,}\qquad\mathcal{H}_{(n\pm1)n} = \frac{\varepsilon}{2} \ee^{\pm \ii\theta},
\end{equation}
where all other elements vanish and we have introduced the momentum class interval $I_k = [-\hbar k/2,\hbar k/2]$, the dimensionless detunings $\delta_n = \left(2n-1\right) q/(\hbar k) + n (n-1)$ and the dimensionless, time-independent Rabi frequency $\varepsilon = \Omega_0/\omega_k$. 
Here, $\Omega_0$ and $\theta$ are modulus and phase of the complex Rabi frequency.
The Rabi frequency is assumed to be constant during the interaction time, which corresponds to box-shaped pulses.
Furthermore, we assume that no additional external potential is present during the interaction.

The interferometric sequence considered here will be driven between the main momentum classes $n=0,1$ and atoms transferred to any of the adjacent classes are regarded as lost.
To quantify this effect, we truncate the system to the momentum classes $n=-1,0,1,2$ and find the matrix  
\begin{equation}
    \mathcal{H} = \frac{1}{2}
    \begin{pmatrix}
		2\delta_{-1} & \varepsilon\ee^{-\ii\theta} & 0 & 0 \\
		\varepsilon\ee^{\ii\theta} & 2\delta_0 & \varepsilon\ee^{-\ii\theta} & 0 \\
		0 & \varepsilon\ee^{\ii\theta} & 2\delta_1 & \varepsilon\ee^{-\ii\theta} \\
		0 & 0 & \varepsilon\ee^{\ii\theta} & 2\delta_2
	\end{pmatrix}.
\end{equation}
\end{subequations}

Our objective is to derive the momentum-class field operators in the Heisenberg picture, making it straightforward to incorporate any initial state and calculate the associated MZI phase uncertainty.
First, we introduce the dimensionless time $\lambda = \omega_k t$, which allows identifying the energy scale on which the adjacent classes are populated.
The equations of motion for the field operators $\hat{\Psi}_{n}$ then take the form
\begin{equation}\label{eq:Heisenberg_eom}
    \ii \partial_\lambda \hat{\Psi}_{m} = \left[ \hat{\Psi}_{m}, \hat{H}/(\hbar \omega_k) \right] = \sum_{n=-1}^{2} \mathcal{H}_{mn} \hat{\Psi}_{n},
\end{equation}
where $m=-1,0,1,2$ and $\mathcal{H}_{mn}$ denotes the $mn$-element of $\mathcal{H}$.
To derive a perturbative solution, we compute the Dyson series up to second order in the coupling strength $\varepsilon$ within an appropriately chosen interaction picture~\cite{SIso2018}, which systematically avoids the appearance of secular terms $\propto\lambda^j$, \cf{} Appendix~\ref{app:Pert-Bragg}.
This procedure yields an accurate description of the main classes $n=0,1$, quantifying the amount of atoms transferred to the adjacent classes up to second order in $\varepsilon$.
We denote the Heisenberg-picture field operators~\footnote{Up to a global phase, the Heisenberg- and interaction-picture field operators differ only by a phase factor, namely $\hat{\Psi}_{n,\mt{H}}(q,\lambda) = \exp[(-1)^n\Delta\omega t/2]\hat{\Psi}_{n}(q,\lambda)$.}
acting on the subspace of the main momentum classes $n=0,1$ by
\begin{equation}\label{eq:time_evo_pulse}
    \hat{\Psi}_{m,\mt{H}}(q,\lambda)\ket{\psi_\mt{in}} = 
      \sum_{n = 0}^1 G_{mn} \hat{\Psi}_n(q,0)\ket{\psi_\mt{in}},
\end{equation}
where $m=0,1$.
This reduced notation is justified because we detect atoms only in the main momentum classes and additionally exclusively consider initial states that populate the same classes, \ie{} $\ket{\psi_\mt{in}} = \ket{\psi}_{0,1}\otimes\ket{0}_{-1,2}$, where $\ket{\psi}_{0,1}$ is an arbitrary state in the product Hilbert space of the main momentum classes, and $\ket{0}_{-1,2}$ denotes the vacuum state in all other. 
Up to a global phase, the time evolution during the laser pulse is given by  
\begin{equation}\label{eq:SIso_solution_Bragg4x4_res_system}
    G(q,\tau, \theta) = 
    \begin{pmatrix}
        T & R \\
        \tilde{R} & \tilde{T} \\
    \end{pmatrix} =
    \begin{pmatrix}
        \tilde{t} & \tilde{r} \\
        -\tilde{r}^* & \tilde{t}^* \\
    \end{pmatrix} 
    \begin{pmatrix}
        \gamma_{00} & \gamma_{01} \\
        \gamma_{10} & \gamma_{11} \\
    \end{pmatrix},
\end{equation}
where $\tau = \varepsilon\lambda = \Omega_0 t$ denotes the pulse area and $\theta$ the initial phase of the light pulse. 
Here, $R$ and $\tilde{R}$ represent the overall reflection coefficients, while $T$ and $\tilde{T}$ denote the transmission coefficients.
Crucially, $G$ is generally non-unitary as it describes diffraction to the adjacent momentum classes, and consequently $T\tilde{T}-R\tilde{R} \leq 1$. 
The time evolution $G$ can be decomposed into an $\mt{SU}(2)$ component and a non-unitary contribution.
Here, $\tilde{t}(q,\tau)$ and $\tilde{r}(q,\tau)$ are the modified transmission and reflection coefficients of the unitary part and the $\gamma_{mn}(q,\tau)$ describe atom loss from the main momentum classes due to parasitic diffraction.
The explicit expressions corresponding to Eq.~\eqref{eq:SIso_solution_Bragg4x4_res_system} are listed in Appendix~\ref{app:Pert-Bragg}.

In the limit of negligible coupling to adjacent classes, one finds $\gamma_{mn} = \delta_{mn}$, and $\tilde{t}$ and $\tilde{r}$ reduce to the usual velocity-selective coupling coefficients~\cite{Szigeti2012-Velocity_selectivity} between the main classes $n=0,1$ [see Eq.~\eqref{eq:0oEps_trans_ref_coef}].
Including the coupling to the adjacent classes up to second order in $\varepsilon$, the transmission and reflection coefficients acquire corrections due to the parasitic diffraction.
Figure~\ref{fig:parabola}(c) shows the time evolution of the atomic momentum distribution during a first-order Bragg pulse for pulse areas $0\leq\tau\leq2\pi$.
The initial distribution is taken to be a Gaussian centered at the resonant momentum $p_0$.
The data are obtained from a numerical calculation well within the quasi-Bragg regime with $\varepsilon = 0.8$.
The density plot clearly illustrates the effect of velocity selectivity.
Especially near $\tau=\pi$, only atoms with momentum close to the resonance are transferred to the first momentum class, whereas atoms with off-resonant momenta predominantly remain in the zeroth class.
Consequently, the beam-splitter pulse becomes imbalanced and the mirror pulse less reflective. 
The figure also shows atom transfer to the adjacent classes.
Both effects are accounted for in our treatment detailed in Appendix~\ref{app:Pert-Bragg}.

\subsection{Interferometric Sequence}
\label{sec:Interferometric-sequence}
\begin{figure}[htb]
	\centering
	\includegraphics{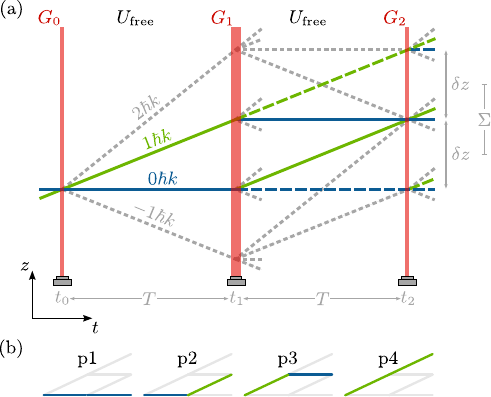}
	\caption{
    (a) Space-time diagram of a Mach--Zehnder atom interferometer driven by first-order Bragg diffraction in absence of external potentials.
    Beam-splitter $G_{0,2}$ and mirror pulses $G_1$ of duration $t_j$ create the main interferometer (solid lines). 
    Between the laser pulses, the atoms propagate freely ($U_\mt{free}$) for an interrogation time $T$.
    The interferometer is driven between the momentum classes $n=0$ (blue) and $n=1$ (green), where the colored dashed paths are introduced by the velocity-selective mirrors, creating additional exits displaced \wrt{} the main exit by $\delta z = \hbar k T/m$, with atomic mass $m$.
    We assume the atomic clouds to be sufficiently localized such that the main exit is contained in $\Sigma = [\delta z/2, 3\delta z/2 ]$. 
    Parasitic diffraction to adjacent momentum classes $n=-1,2$ populates the spurious paths (dashed gray).
    Spurious paths coupling into the interferometer exit are not populated to the considered order of perturbation.
    (b) Legend for the paths p$j$ contributing to the interference signal.
	}
	\label{fig:MZI_w_sp_paths}
\end{figure}

Figure~\ref{fig:MZI_w_sp_paths}(a) depicts the space-time graph of an MZI driven by first-order Bragg diffraction in the absence of acceleration or any additional external potential. 
The main interferometer consists of the solid paths (blue, green), whereas the dashed paths are introduced by velocity selectivity and parasitic diffraction, affecting the phase uncertainty of the MZI.
Velocity selectivity degrades the phase sensitivity in two ways.
First, the reduced reflectivity of the mirror pulse creates two additional exits that are spatially displaced, and whose signal deteriorates the signal-to-noise ratio if they cannot be resolved~\cite{Jenewein2022-Spurious_paths_Bragg_AI}.
Second, imperfect beam-splitter pulses lead to unequal population of the arms of the main interferometer.
In addition, diffraction to the adjacent classes $n=-1,2$ gives rise to further paths shown as dashed gray lines in Fig.~\ref{fig:MZI_w_sp_paths}(a). 
Atoms that are diffracted onto these paths are lost from the interferometer sequence.
This approximation is justified because the population on the four gray paths that couple into exits is negligible to the perturbative order considered here.
Therefore, in the following we restrict the analysis to the colored paths p$j$ indicated in Fig.~\ref{fig:MZI_w_sp_paths}(b).

With this restriction, the MZI can be effectively described~\cite{Günther2024-squeezingenhancementlossymultipath} within the subspace spanned by the relevant classes $n=0,1$.
The complete time evolution is then obtained as a sequence of matrix transformations for the respective light pulses given by  Eq.~\eqref{eq:SIso_solution_Bragg4x4_res_system} and the free propagation in between the pulses.
The Heisenberg-picture field operators of the relevant momentum classes $n=0,1$ at the final dimensionless time $\lambda_\mt{f}=\omega_k(t_0+t_1+t_2+2T)$, where $t_j$ are the pulse durations and $T$ is the interrogation time, are given by
\begin{equation}\label{eq:Heisenberg_pic_field_ops}
    \hat{\Psi}_{m,\mt{H}}(q,\lambda_\mt{f})\ket{\psi_\mt{in}} = \sum_{n = 0}^1 M_{mn} \hat{\Psi}_n(q,0)\ket{\psi_\mt{in}}.
\end{equation}
The overall time evolution matrix is given by
\begin{equation}\label{eq:MZI_trafo}
    M = G_2 U_\mt{free} G_1 U_\mt{free} G_0,
\end{equation}
where $G_{0,2} = G(q,\pi/2,\theta_{0,2})$ represent the beam-splitter pulses and $G_{1} = G(q,\pi,\theta_1)$ the mirror pulse, which primarily differ in their pulse areas.
Here, $\theta_j$ denotes the initial phase of the respective light field corresponding to the argument of the complex Rabi frequency.
Following Eq.~\eqref{eq:SIso_solution_Bragg4x4_res_system}, the transmission and reflection coefficients entering the matrices $G_j$ are denoted by $\tilde{t}_j$, $\tilde{r}_j$, $T_j$, $R_j$, $\tilde{T}_j$ and $\tilde{R}_j$. 
Generally, we assume identical laser intensities for all pulses, such that the mirror pulse is realized by doubling the pulse duration compared to the beam splitters, \ie{} $t_1 = 2t_{0,2} = \pi/\Omega_0$, but we are not limited to this configuration.
The free propagation is given by 
\begin{equation}
	U_\mt{free} = \mt{diag}\,\Big(\exp[\ii \Delta\varphi_{\mt{free}}/2],\exp[-\ii\Delta\varphi_{\mt{free}}/2]\Big)
\end{equation}
up to a global phase, where the relative phase imprinted onto the momentum classes is given by $\Delta\varphi_{\mt{free}} = [\nu_k(q) + \Delta \omega] T$.
Here, we used the resonance condition for first-order Bragg diffraction, \ie{} $\Delta\omega = \nu_k(p_0) + \omega_k$.
Note that the sequence in Eq.~\eqref{eq:Heisenberg_pic_field_ops} provides a valid description of the interferometer dynamics only if the atomic density is sufficiently low to suppress nonlinear interactions after the first beam splitter~\cite{Feldmann2023-optimalsqueezinghighprecisionatom}.

\subsection{Observables and Initial States}
\label{sec:Pseudo-ang-mom}
To describe the propagation of possibly entangled atomic states through the interferometer, we calculate the observable measured at the MZI output in the Heisenberg picture using Eq.~\eqref{eq:Heisenberg_pic_field_ops}.
Throughout this work, we choose the atom-number difference between the main momentum classes $n=0,1$ as observable.
In the framework of pseudo-spin angular momentum~\cite{schwinger2015angular, Pezze2018-Qrev}, this quantity corresponds to the angular momentum $\hat{J}_3$ along the quantization axis $x_3$.
The Heisenberg-picture observable is given by
\begin{equation}\label{eq:J3_observable}
	\hat{J}_{3,\mt{H}} = \frac{1}{2}\int_{I_k}\!\dd q\,\bigl(\hat{\Psi}_{1, \mt{H}}^\dag\hat{\Psi}_{1,\mt{H}}^{\phantom{\dag}}-\hat{\Psi}_{0,\mt{H}}^\dag\hat{\Psi}_{0,\mt{H}}^{\phantom{\dag}}\bigr),
\end{equation}
where $\hat{\Psi}_{n,\mt{H}}$ in turn is given by Eq.~\eqref{eq:Heisenberg_pic_field_ops}.
This choice enables us to relate the observable to the initial angular-momentum quadratures and leads to intuitive and compact expressions for both the interference signal and the phase uncertainty~\cite{Günther2024-squeezingenhancementlossymultipath}, as shown below.

Generally, we assume that all atoms have the same initial momentum distribution $|\varphi_0(q)|^2$ with support $I_k$, centered at the respective central momentum of each class.
This assumption corresponds to a single-mode atomic source, such as a Bose--Einstein condensate, and describes the common wavefunction of each atom rather than an incoherent distribution over different external modes.
Under this assumption, the atoms in the input of the interferometer can be described by the multi-mode operators
\begin{equation}
    \hat{a}_j = \int_{I_k}\!\dd q\,\varphi_0^*(q)\hat{\Psi}_j(q),
\end{equation}
where $\varphi_0$ is square integrable.
The multi-mode operators fulfill the discrete bosonic commutation relations $\big[\hat{a}_i^{\phantom{\dag}},\hat{a}_j^\dag\big] = \delta_{ij}\id$ and $\big[\hat{a}_i,\hat{a}_j\big] = \hat{0}$.
We describe initial atomic states utilizing the multi-mode operators within the framework of pseudo-angular momentum and introduce the initial angular momentum operators
\begin{equation}
\label{eq:J3_observable}
	\hat{S}_\ell = \big(\hat{a}_1^\dag, \hat{a}_0^\dag\big) \frac{\sigma_\ell}{2}
    \begin{pmatrix}
        \hat{a}_1\\
        \hat{a}_0
    \end{pmatrix},
\end{equation}
where $\sigma_\ell$ are the Pauli matrices and $\ell=1,2,3$ denotes the quantization direction $x_\ell$.
An important advantage of this formalism is that we can express the interference signal $\langle\hat{J}_{3,\mt{H}}\rangle$ as a function of the angular momentum quadratures $\langle \hat{S}_\ell \rangle$ associated with the initial state. 
In this notation, the signal's variance $\Delta J_{3,\mt{H}}^2$ is a function of the covariances between the initial angular momentum operators $\operatorname{Cov}[\hat{S}_i,\hat{S}_j] = \langle\hat{S}_i\hat{S}_j + \hat{S}_j\hat{S}_i\rangle/2 - \langle\hat{S}_i\rangle\langle\hat{S}_j\rangle$ and the variances $\Delta S_i^2 = \operatorname{Var}[\hat{S}_i] = \operatorname{Cov}[\hat{S}_i,\hat{S}_i]$.

\section{Removing the Quasi-Incoherent Background}
\label{sec:rm_background}
In this section, we exclusively consider the effect of velocity selectivity and neglect coupling to the adjacent momentum classes $n=-1,2$. 
This situation corresponds to the MZI shown in Fig.~\ref{fig:MZI_w_sp_paths}(a) without the dashed gray paths.
In this case, the equation of motion [Eq.~\eqref{eq:Heisenberg_eom}] is analytically solvable and the time evolution $G$, given by Eq.~\eqref{eq:SIso_solution_Bragg4x4_res_system}, simplifies accordingly.
Since there is no atom loss from the main classes, we have $\gamma_{ij} = \delta_{ij}$ and the transmission and reflection coefficients reduce to the standard expressions 
\begin{subequations}\label{eq:0oEps_trans_ref_coef}
    \begin{align}
        \tilde{t}_j &\rightarrow \ee^{\ii \tilde{\varphi}_j}\left(\cos\frac{f \tau_j}{2} + \ii\frac{v\tau_j}{2}\sinc\frac{f \tau_j}{2}\right), \\
    	\tilde{r}_j &\rightarrow -\ii\ee^{\ii(\tilde{\varphi}_j - \theta_j)}\frac{\tau_j}{2}\sinc\frac{f \tau_j}{2}
    \end{align}
\end{subequations}
for detuned Rabi oscillations, where $\tilde{\varphi}_j = \Delta\omega t_j/2$ is the dynamic differential phase imprinted by the counterpropagating lasers, $\tau_j = \Omega_0 t_j$ the pulse area and $\theta_j$ the initial phase of the $j$th light pulse~\cite{Szigeti2012-Velocity_selectivity}.
We have introduced the dimensionless effective Rabi frequency $f = \sqrt{1+v^2}$ and the dimensionless Doppler detuning $v= \nu_k/\Omega_0$.
 
The time evolution describing the MZI is given by the sequence of Eq.~\eqref{eq:MZI_trafo}.
For this section, we consider an initial Fock state with all $N$ atoms in momentum class $n=0$, \ie{} $\ket{\psi_\mt{in}} = \bigl(\hat{a}_0^\dag\bigr)^{N}\ket{0}/\sqrt{N!}$, where we assume that the atoms have a Gaussian momentum distribution, with standard deviation $\sigma_q$, that is centered at the resonant momentum $p_0$.
This state is not entangled and only populates a single input of the interferometer, such that it is feasible to describe the dynamics within first quantization.
Accordingly, the results presented in this section are also readily derived within a first-quantized single-particle calculation.
Because the Fock state does not exhibit first-order coherence, namely, $\langle \hat{a}_1^\dag\hat{a}_0^{\phantom{\dag}} \rangle = 0$, the interference signal takes the form
\begin{equation}\label{eq:Int_signal_only_VS}
	\langle\hat{J}_{3,\mt{H}}\rangle 
    = \mathscr{A}_0 \langle\hat{S}_3\rangle 
    = - \mathscr{A}_0N/2,
\end{equation}
where $\mathscr{A}_0$ denotes the population-difference contribution. 
It essentially represents the interference signal, weighted by the initial momentum distribution of the atoms 
\begin{equation}\label{eq:wed_int_contrib}
	\mathscr{A}_0(\phi) = \int_{I_k}\!\dd q\, |\varphi_0|^2 \big( |M_\mt{00}|^2 - |M_\mt{10}|^2 \big),
\end{equation}
with the matrix element $M_{i0}$ describing atoms entering the interferometer in class $n=0$ and exiting in $n=i$. 
The population-difference contribution depends on the interferometer phase
\begin{equation}\label{eq:interferometer_phase}
	\phi = \theta_0 - 2\theta_1 + \theta_2 - \Delta\omega\pi/ (2\Omega_0),
\end{equation}
which consists of the discrete second derivative of the initial laser phases and the total dynamic relative phase $\Delta\omega\pi/ (2\Omega_0)$ accumulated during the light pulses.
The matrix elements can be decomposed into contributions from the individual paths shown in Fig.~\ref{fig:MZI_w_sp_paths}(b) and are given by
\begin{equation}\label{eq:U_00}
    M_{i0} = M_\mt{p1}^{(i)} + \ee^{-\ii \Delta\varphi_{\mt{free}}} \left(M_\mt{p2}^{(i)} +  M_\mt{p3}^{(i)}\right) + \ee^{-2\ii \Delta\varphi_{\mt{free}}} M_\mt{p4}^{(i)},
\end{equation}
up to a global phase.
The phases $n\Delta\varphi_{\mt{free}}$ with $n=1,2$ contain the Doppler phases $n\nu_k T$, which act as a spatial displacement of the respective path, displacing the atomic clouds by $n \delta z = n \hbar k T/m$.
The exit positions can be directly associated with the paths depicted in Fig.~\ref{fig:MZI_w_sp_paths}(a).
The schematic also provides intuition for the explicit form of the path contributions $M_{\mt{p}j}^{(i)}$, which each corresponding to a sequence of transmission and reflection coefficients that are given in Table~\ref{tab:Ms_expressions}.
\bgroup
\def\arraystretch{1.6}
\begin{table}[htb]
    \caption{
    Contributions $M_{\mt{p}j}^{(i)}$ to the matrix elements $M_{i0}$ given by Eq.~\eqref{eq:U_00}.
    They correspond to sequences of reflection and transmission coefficients in accordance with the respective paths in Fig.~\ref{fig:MZI_w_sp_paths}.
    } 
    \centering
    \begin{tabular}{l c c c c}
        \hline\hline
        $i\, \backslash\, j$& 1 & 2 & 3 & 4 \\ \hline
        0 & $\tilde{t}_{2}^{\phantom{*}} \tilde{t}_{1}^{\phantom{*}} \tilde{t}_{0}^{\phantom{*}}$ & $- \tilde{r}_{2}^{\phantom{*}} \tilde{r}_{1}^* \tilde{t}_{0}^{\phantom{*}}$ & $- \tilde{t}_{2}^{\phantom{*}} \tilde{r}_{1}^{\phantom{*}} \tilde{r}_{0}^*$ & $- \tilde{r}_{2}^{\phantom{*}} \tilde{t}_{1}^* \tilde{r}_{0}^*$ \\
        1 & $-\tilde{r}_{2}^* \tilde{t}_{1}^{\phantom{*}} \tilde{t}_{0}^{\phantom{*}}$ & $-\tilde{t}_{2}^* \tilde{r}_{1}^* \tilde{t}_{0}^{\phantom{*}}$ & $\tilde{r}_{2}^* \tilde{r}_{1}^{\phantom{*}} \tilde{r}_{0}^*$ & $-\tilde{t}_{2}^* \tilde{t}_{1}^* \tilde{r}_{0}^*$ \\
        \hline\hline
    \end{tabular}
    \label{tab:Ms_expressions}
\end{table}
\egroup

\begin{figure}[htb]
	\centering
	\includegraphics{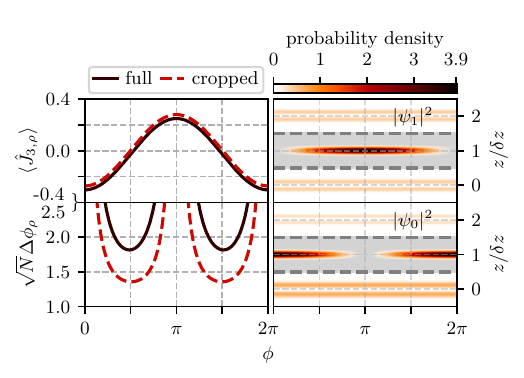}
    \caption{
        (Top left panel) Interference signals $\langle\hat{J}_{3,\rho}\rangle$ obtained from scanning the interferometer phase $\phi$ including ($\rho=\mt{H}$, full) and without ($\rho=\Sigma$, cropped) the quasi-incoherent background signal for $N =1$.
        The atom has initially a Gaussian momentum distribution, with standard deviation $\sigma_q=0.05\hbar k$, centered at resonant momentum $p_0$. 
        The Rabi frequency is set to $\Omega_0 = 0.1\omega_k$ and the interrogation time to $T=10^3/\Omega_0$.
        The cropped signal is slightly more centered around zero than the full signal, implying a decreased phase uncertainty (bottom left panel).
        (Right) Spatial distribution of an atom directly after the MZI sequence in the momentum class $n=1$ (top panel, $|\psi_1|^2$) and $n=0$ (bottom panel, $|\psi_0|^2$). 
        The gray-shaded region $\Sigma = [\delta/2, 3\delta/2]$ contains the interferometer's main exit, where clear interference fringes between the momenta are visible.
        Outside $\Sigma$, the displaced exits show no modulation with the interferometer phase (due to negligible overlap between exits), thereby contributing a quasi-incoherent background to the interferometer signal. 
        The outer exits exhibit a spatial two-peaked structure caused by the velocity-selective mirror pulse. 
        Only atoms with (near-)resonant momenta are diffracted into the main exit, which depopulates the spatial center, leaving only off-resonant atoms to remain that create the two-peaked structure.
	}
	\label{fig:full_vs_cutout_signal}
\end{figure}

The squared matrix elements $|M_{i0}|^2$ entering the population-difference contribution in Eq.~\eqref{eq:wed_int_contrib} contain all pairwise overlaps $M_{\mt{p}j}^{(i)*}M_{\mt{p}k}^{(i)}$ between two paths p$j$ and p$k$.
We restrict our analysis to a regime in which the three atomic clouds in the exit port of the interferometer are sufficiently localized such that they do not overlap after the MZI sequence. 
Under this condition, interference of distinct paths involving the outer paths p1 and p4 can be neglected in Eq.~\eqref{eq:wed_int_contrib}.
This amounts to negligible overlaps $\int\!\dd q\, |\varphi_0|^2 \exp(\pm\ii n \nu_k T) M_{\mt{p}i}^{(\ell)*}M_{\mt{p}j}^{(\ell)}$ for Doppler phase factors $\exp(\pm\ii n \nu_k T)$, where $n\neq0$.
Although the outer paths do not interfere with any other contribution, their population in the respective momentum class still contributes to the detected signal.
The populations $|M_{\mt{p}1,\mt{p}4}^{(\ell)}|^2$ therefore act as a quasi-incoherent background, decreasing the overall contrast of the signal~\cite{Jenewein2022-Spurious_paths_Bragg_AI}.
Indeed, after neglecting interference involving the outer paths, the interference signal becomes 
\begin{equation}\label{eq:Int_signal_only_VS}
	\langle\hat{J}_{3,\mt{H}}\rangle 
    = \frac{N}{2}(\mathscr{O}_\mt{H} - \mathcal{J}\cos\phi),
\end{equation}
where $\phi$ denotes the interferometer phase defined in Eq.~\eqref{eq:interferometer_phase}.
The signal scales with the initial number $N$ of input atoms and is characterized by the amplitude $\mathcal{J}$ and an offset $\mathscr{O}_\mt{H}$.
The latter is introduced by velocity selectivity and reads
\begin{equation}\label{eq:offset_only_VS}
   \mathscr{O}_\mt{H} = -\int_{I_k}\!\dd q\, \frac{|\varphi_0|^2}{f^{6}}\left(v^{2} + \cos\!\frac{\pi f}{2}\right)^{2}\left[v^{2} + \cos(\pi f)\right],
\end{equation}
which contains the quasi-incoherent background associated with the outer paths p1 and p4.
Velocity selectivity also reduces the signal's amplitude, which is given by
\begin{equation}
    \mathcal{J} = \int_{I_k}\!\dd q\, \frac{|\varphi_0|^2}{f^{6}}\left(1 + v^{2}\sec^{2}\!\frac{\pi f}{4}\right)\sin^{4}\!\frac{\pi f}{2}.
\end{equation}
Figure~\ref{fig:full_vs_cutout_signal} (top left panel) displays a full fringe scan of the interference signal from Eq.~\eqref{eq:Int_signal_only_VS}, shown as the black line.
One observes that the sinusoidal signal is not symmetrically centered around $\langle\hat{J}_3\rangle = 0$. 
This effect is caused by the offset $\mathscr{O}_\mt{H}$ given by Eq.~\eqref{eq:offset_only_VS}.
In addition, the signal's amplitude is reduced by velocity selectivity, degrading the associated phase sensitivity.
For reference, in the absence of velocity selectivity the interference signal is $\langle\hat{J}_3\rangle =-N/2\cos\phi$, corresponding to a vanishing offset $\mathscr{O}_\mt{H}=0$ and unit amplitude $\mathcal{J} = 1$.

Fourier transforming the momentum-mode function $\varphi_0(q)M_{i0}(q)$ of atoms exiting the interferometer in class $n=i$ yields the corresponding position distribution $|\psi_i(z,\phi)|^2$, shown in Fig.~\ref{fig:full_vs_cutout_signal} for class $n=0$ (bottom right panel) and $n=1$ (top right panel).
The central contribution, created by paths p2 and p3, exhibits the interference fringe as an oscillation of the population between $n=1$ and $n=0$. 
While we can distinguish different momentum classes in our formalism, the spatial distributions of atoms in these two momentum classes are at the same position.
However, because the atoms carry different momenta, the position distributions of the two classes can be resolved after a suitable time of flight.

The two displaced exit ports created by paths p1 and p4 are phase insensitive and only contribute as a quasi-incoherent background to the signal because of their negligible spatial overlap with the main paths p2 and p3.
Their spatial distributions clearly reveal the effect of velocity selectivity.
The two-peaked structure contains off-resonant atoms that are not addressed by the mirror pulse, while the atoms close to the center of the structure have been resonantly coupled into the main exit.
This quasi-incoherent background degrades the signal-to-noise ratio by contributing to the offset in Eq.~\eqref{eq:Int_signal_only_VS}. 
We discuss its effect on the phase sensitivity below.

Motivated by the above observations, we exclude the quasi-incoherent background by only detecting the atoms in the main exit, which is expected to reduce the signal's offset and thus enhance the phase sensitivity.
This procedure is facilitated by projecting the observable onto the spatial interval $\Sigma = [\delta z/2, 3\delta z/2 ]$ indicated by the  gray shaded region in Fig.~\ref{fig:full_vs_cutout_signal}, thereby removing the  contributions from the outer exits centered at $z=0$ and $z=2 \delta z$.
This procedure relies on the assumption that the atomic clouds are sufficiently localized and separated.
The projected observable is given by
\begin{equation}\label{eq:sector_projector}
	\hat{J}_{3,\Sigma} = \frac{1}{2}\int_{\Sigma}\!\dd z\, \left[\hat{\Psi}_{1,\mt{H}}^\dag(z) \hat{\Psi}_{1,\mt{H}}^{\phantom{\dag}}(z) - \hat{\Psi}_{0,\mt{H}}^\dag(z) \hat{\Psi}_{0,\mt{H}}^{\phantom{\dag}}(z)\right], 
\end{equation}
where we have introduced the position-space field operators 
\begin{equation}
	\hat{\Psi}_{n,\mt{H}}(z) = \frac{1}{\sqrt{2\pi\hbar}}\int_{I_k}\!\dd q\, \ee^{\ii q z / \hbar}\hat{\Psi}_{n,\mt{H}}(q,\lambda_\mt{f}).
\end{equation}
The resulting cropped interference signal takes the same form as the full signal
\begin{equation}\label{eq:Int_signal_only_VS_cropped}
	\langle\hat{J}_{3,\Sigma}\rangle 
    = \frac{N}{2}(\mathscr{O}_\Sigma - \mathcal{J}\cos\phi),
\end{equation}
but with a modified offset $\mathscr{O}_\Sigma$ that no longer contains the quasi-incoherent background and is explicitly given by
\begin{equation}
   \mathscr{O}_\Sigma = \int_{I_k}\!\dd q\, \frac{|\varphi_0|^2}{f^{6}}\left(v^{2} + \cos\!\frac{\pi f}{2}\right)^{2}\sin^{2}\!\frac{\pi f}{2}.
\end{equation}
As a consequence, the offset is reduced as illustrated in Fig.~\ref{fig:full_vs_cutout_signal} (top left panel).
However, it does not vanish entirely due to the population imbalance on the main paths p2 and p3 introduced by the velocity-selective beam-splitter pulses.
Technically, both $\langle\hat{J}_{3,\Sigma}\rangle$ and the variance $\Delta J_{3,\Sigma}^2$ can be derived from the corresponding full-signal quantities by excluding the contribution transmitted at the mirror pulse, \ie{} by setting $T_1  = 0 = \tilde{T}_1$.

The phase uncertainty is calculated via Gaussian uncertainty propagation
\begin{equation}\label{eq:PS_gauss_err_prop}
	\Delta\phi_\rho^2 = \frac{\Delta J_{3,\rho}^2}{\bigl(\partial_\phi\langle\hat{J}_{3,\rho}\rangle\bigr)^2},
\end{equation}
with $\rho=\Sigma,\mt{H}$ denoting the cropped and full measurements.
The variance can be computed for an arbitrary offset and amplitude, yielding
\begin{equation}
    \Delta J_{3,\rho}^2 = \frac{N}{4}\left[ \eta_\rho - (\mathscr{O}_\rho - \mathcal{J}\cos\phi)^2 \right],
\end{equation}
where $\eta_\rho$ is the respective fraction of detected atoms, \ie{} $\eta_\mt{H} = 1$, and $\eta_\Sigma = \int\!\dd q\,|\varphi_0 \tilde{r}_{1}|^2$ being the weighted mirror reflectivity.
The resulting phase uncertainty is
\begin{equation}\label{eq:PS_only_VS}
	\Delta\phi_\rho^2 
    = \frac{1}{N}\left(\frac{\eta_\rho-\mathscr{O}_\rho^2}{\mathcal{J}^2\sin^2\!\phi} + \frac{2\mathscr{O}_\rho\cos\phi}{\mathcal{J}\sin^2\!\phi} - \cot^2\!\phi\right),
\end{equation}
which directly shows that the sensitivity is bound by SNL.
As shown in Fig.~\ref{fig:full_vs_cutout_signal} (bottom left panel), the reduced offset in the cropped signal leads to a significant improvement of the phase sensitivity.
Although the phase uncertainty of the cropped signal approaches SNL, namely, $\Delta\phi^2 = 1/N$, it does not fully saturate the bound due to atoms lost to the quasi-incoherent background and the residual population imbalance on the paths.

To resolve the spatial distribution of the atoms and discriminate between the two momentum classes after the interferometer, detection must be performed after an intermediate time of flight $T_\text{tof}$. 
Immediately after the interferometer sequence, the spatial distributions of the two classes overlap, whereas in the far-field limit, the spatial separation fully vanishes. 
We emphasize that the procedure of cropping the background signal can only be performed if the atoms are sufficiently spatially localized and separated.
Overlapping atomic clouds from different exit ports introduce a beating in the signal that cannot be removed by cropping and further degrades the phase sensitivity~\cite{Jenewein2022-Spurious_paths_Bragg_AI, Kirsten-Siemss2023}.
The cropping procedure requires the wave packet width after the total expansion time $\sigma_z^2(2T+T_\text{tof}) \ll \delta z^2$, where we have $\sigma_z^2(t) = \sigma_z^2(0) + (\sigma_q t /m)^2$ for Gaussian wave packets.
In addition, $T_\text{tof}$  must be sufficiently long to separate neighboring momentum classes including the displaced exit ports. 
While these are displaced by $\pm \delta z = \pm \hbar kT/m$ after the interferometer, the adjacent momentum classes separate ballistically during the time of flight by $\hbar k T_\text{tof}/m $.
Thus, choosing $T_\text{tof}\gtrsim 3T$ provides sufficient spatial separation to resolve the momentum classes and spurious exit ports.
As a representative estimate, assuming a cloud of \rb{} atoms with $\sigma_q=0.005 \hbar k $ and a minimum-uncertainty state with $\sigma_z(0)= \hbar / (2\sigma_q)$ yields $\sigma_z(5T)/ \delta z< 1/5$ for $T\gtrsim 2.7$\,ms assuming a recoil frequency of $2\pi \times 15.1 \, \text{kHz}$.
Larger initial cloud sizes require correspondingly longer interrogation times.

In the following section, we present a more general expression for the phase sensitivity that accounts for diffraction into adjacent momentum classes and incorporates the description of entangled input states.

\section{Combining Loss due to Velocity Selectivity and Higher Diffraction Orders}
\label{sec:PS_including_VS_and_PD}
To incorporate the coupling to the adjacent momentum classes $n=-1,2$, we use the perturbative solution for the first-order Bragg pulses from Eq.~\eqref{eq:time_evo_pulse} and the interferometer time evolution given by Eq.~\eqref{eq:Heisenberg_pic_field_ops}.
Since we include second-order corrections in $\varepsilon$ to the transmission and reflection coefficients of the $j$th pulse, they are given by the full quantities $\tilde{T}_j$, $T_j$, $\tilde{R}_j$ and $R_j$ listed in Appendix~\ref{asec:pert_sol_bragg_diff}.
In accordance with our findings in Sec.~\ref{sec:rm_background}, we employ the same detection scheme, \ie{} only detecting atoms in the interferometer's main exit and resolving the momentum classes after a suitably long time of flight.
We further assume that the three atomic clouds in the interferometer exits depicted in Fig.~\ref{fig:MZI_w_sp_paths}(a) do not overlap.

Our goal is to determine the phase uncertainty $\Delta\phi$ in dependence on the initial angular momentum properties.
To this end, we calculate the signal $\langle \hat{J}_\mt{3,H} \rangle$, where the Heisenberg-picture field operators $\hat{\Psi}_{n,\mt{H}}$ are described by the time evolution given by Eq.~\eqref{eq:Heisenberg_pic_field_ops}. 
Additionally, we set $\tilde{T}_1 = 0 = T_1$ to exclude the quasi-incoherent background corresponding to the cropping in position space discussed above.
Both, the interference signal and its variance depend on the overlaps $M_{ij}^*M_{kl}$ of the matrix elements of the time evolution from Eq.~\eqref{eq:MZI_trafo}.
These matrix elements are determined up to second order in the dimensionless Rabi frequency $\varepsilon$ using a Dyson expansion~\footnote{In our perturbative treatment, the matrix elements contain $\varepsilon$-dependent amplitudes as well as trigonometric functions with $\varepsilon$-dependent frequencies. 
To obtain the matrix elements to second-order accuracy in $\varepsilon$, we apply power counting to the amplitudes and only keep terms up to second order in $\varepsilon$, while preserving trigonometric functions, as they do not contribute to the power counting. 
Expanding them would reintroduce secular terms that are carefully avoided~\cite{SIso2018} within the perturbative treatment of Appendix~\ref{app:Pert-Bragg}.}.

From the signal's derivative $\partial_\phi \langle \hat{J}_\mt{3,H} \rangle$ and variance $\Delta J_\mt{3,H}^2$, we obtain the phase uncertainty via Eq.~\eqref{eq:PS_gauss_err_prop}.
We evaluate the latter at interferometer phase $\phi = 0$ and the first beam-splitter phase $\theta_0 = 0$, since this is the working point of an ideal MZI, unaffected by velocity selectivity and parasitic diffraction, for the entangled input state we will consider below.
This choice gives the phase uncertainty 
\begin{equation}\label{eq:PS_4x4_Bragg_MZI}
    \Delta\phi^2|_{\phi,\theta_0=0} = \frac{\operatorname{Var}\left[|\mathscr{A}_{10}|\hat{S}_{\varphi_{10}} + \mathscr{A}_{0}\hat{S}_3\right] + \frac{V_\mt{M}}{4} \langle\hat{N}\rangle  + \frac{\mathscr{R}_{10}}{2} \langle\hat{S}_{1}\rangle}{\left(\left|\mathscr{A}_{10}'\right|\langle\hat{S}_{\varphi^\prime_{10}}\rangle 
    - \frac{\mathscr{A}_{0}'}{2}\langle\hat{N}\rangle\right)^{2}},
\end{equation}
in terms of the initial angular-momentum properties $\langle\hat{S}_i\rangle$ and $\operatorname{Cov}[\hat{S}_i,\hat{S}_j]$.
The quantities $\mathscr{A}_0$ and $\mathscr{A}_{10}$ are the population-difference and first-order-coherence contributions to the interference signal, respectively.
They correspond to overlaps proportional to  $\hat{\Psi}_n^\dag\hat{\Psi}_n^{\phantom{\dag}}$ and $\hat{\Psi}_m^\dag\hat{\Psi}_n^{\phantom{\dag}}$ (with $n\neq m$), weighted by the initial momentum distribution $|\varphi_0|^2$.
The population-difference contribution was already introduced in the simplified case above. 
We further define the angular-momentum operator $\hat{S}_\varphi = \cos\varphi\hat{S}_1 - \sin\varphi\hat{S}_2$ describing a rotation by an angle $\varphi$ in the $x_1$-$x_2$ plane. 
The phases $\varphi_{10}$ and $\varphi_{10}'$ are the complex arguments of $\mathscr{A}_{10}$ and its derivative $\mathscr{A}_{10}'$ \wrt{} the interferometer phase $\phi$ (both evaluated at $\phi = 0 = \theta_0$).
We denote by $\hat{N}$ the total initial particle operator, with $N = \langle\hat{N}\rangle$ being the initial atom number.
The term $V_\mt{M}$ describes the vacuum noise coupling into the empty input ports of the non-ideal mirror pulses, thereby introducing a shot-noise-limited term.
Similarly, the quantity $\mathscr{R}_{10}$ introduces a shot-noise-limited term depending on first-order coherence measured by $\langle \hat{S}_1 \rangle$.
The remaining term in Eq.~\eqref{eq:PS_4x4_Bragg_MZI} contains the initial angular-momentum expectation values and covariances. 
This contribution is the one enabling quantum enhancement of the phase sensitivity as we will demonstrate below.
The explicit expressions of the script quantities appearing in Eq.~\eqref{eq:PS_4x4_Bragg_MZI} are listed in Table~\ref{tab:scr_quantities}.
\bgroup
\def\arraystretch{1.6}
\begin{table}[b]
    \centering
    \caption{Script quantities appearing in the phase uncertainty from Eq.~\eqref{eq:PS_4x4_Bragg_MZI}. Interferometric contributions $\mathscr{A}_0$ and $\mathscr{A}_{10}$, their derivatives $\mathscr{A}_0'$ and $\mathscr{A}_{10}'$ \wrt{} the interferometer phase $\phi$, the detected fraction of atoms $\mathscr{R}_0$ entering in class $n=0$ and the shot-noise inducing term $\mathscr{R}_{10}$ are given by the corresponding weighted integrand $\bullet$ evaluated at phases $\phi=0=\theta_0$, \ie{} $\int_{I_k}\!\dd q\, \bullet|_{\phi,\theta_0=0} |\varphi_0|^2  + \mathcal{O}(\varepsilon^3)$.
    The matrix elements $M_{ij}$ describe the MZI time evolution with the outer exits (paths p1 and p4 in Fig.~\ref{fig:MZI_w_sp_paths}) already excluded. 
    The script quantities are accurate to second order in $\varepsilon$, consistent with the second-order calculation of $M_{ij}$.}
    \begin{tabular}{l c}
        \hline\hline
        Symbol & Integrand \\ \hline
        $\mathscr{A}_{0}$ &   $ |M_{00}|^2 -  |M_{10}|^2$\\ 
        $\mathscr{A}_{10}$ & $M_\mt{11}^*M_\mt{10}^{\phantom{*}} - M_\mt{01}^*M_\mt{00}^{\phantom{*}}$ \\ 
        $\mathscr{A}_{0}'$ & $\partial_\phi \big(|M_{00}|^2 -  |M_{10}|^2\big)$\\ 
        $\mathscr{A}_{10}'$ & $ \partial_\phi \big(M_\mt{11}^*M_\mt{10}^{\phantom{*}} - M_\mt{01}^*M_\mt{00}^{\phantom{*}}\big)$\\       
        $\mathscr{R}_{0}$ & $|M_{00}|^2 +  |M_{10}|^2$\\ 
        $\mathscr{R}_{10}$ & $M_\mt{11}^{*} M_\mt{10}^{\phantom{*}} +  M_\mt{01}^{*} M_\mt{00}^{\phantom{*}}$\\ 
         \hline\hline
    \end{tabular}
    \begin{tabular}{c}
        \hline\hline
        Matrix elements \\ \hline
        $M_{00} = R_2 \tilde{R}_1 T_0 + T_2 R_1 \tilde{R}_0 $ \\ 
        $M_{01} = R_2 \tilde{R}_1 R_0 + T_2 R_1 \tilde{T}_0 $\\ 
        $M_{11} = \tilde{T}_2 \tilde{R}_1 R_0 + \tilde{R}_2 R_1 \tilde{T}_0 $ \\ 
        $M_{10} = \tilde{T}_2 \tilde{R}_1 T_0 + \tilde{R}_2 R_1 \tilde{R}_0 $ \\ \hline\hline
        \multicolumn{1}{c}{}\\\multicolumn{1}{c}{}\\
    \end{tabular}
    \label{tab:scr_quantities}
\end{table}
\egroup
The vacuum contribution $V_\mt{M}$ is given by
\begin{equation}
    V_\mt{M} = \mathscr{R}_0 - \mathscr{A}_{0}^2 - |\mathscr{A}_{10}|^2,
\end{equation}
where $\mathscr{R}_0$ denotes the fraction of detected atoms that enter the interferometer in momentum class $n=0$.

To gain intuition, we consider the case of an ideal MZI, where coupling to the adjacent momentum classes is neglected and velocity selectivity excluded. 
The latter corresponds to the delta-pulse approximation~\cite{Schleich2013, Funai2019}, \ie{} infinitely short pulses with fixed pulse area.
In this limit, the interferometric contributions are
\begin{equation}\label{eq:A_0_A_01_delta_puslse_apprx}
	\mathscr{A}_0(\phi) = \cos\phi, \qquad
	\mathscr{A}_{10}(\phi,\theta_0) = \ee^{\ii\theta_0}\sin{\phi},
\end{equation}
and the quantities introducing shot noise vanish, \ie{} $V_\mt{M} = 0 = \mathscr{R}_{10}$.
Notably, the interferometric contributions $\mathscr{A}_0$ and $\mathscr{A}_{10}$ are independent of the momentum distribution, as there is no velocity selectivity.
The phase uncertainty at $\phi = 0 = \theta_0$ according to Eq.~\eqref{eq:PS_4x4_Bragg_MZI} reads 
\begin{equation}\label{eq:PS_id_MZI}
    \Delta\phi^2 = \frac{\Delta S_3^2}{\langle\hat{S}_1\rangle^2},
\end{equation}
which corresponds to the ratio of the variance of the angular momentum in the $x_3$ direction and the squared projection onto the $x_1$ direction.
Hence, reducing the phase uncertainty requires an initial atomic state that is strongly polarized along $x_1$, while exhibiting a small variance along $x_3$.
Quantum enhancement in the MZI can therefore be achieved by employing pseudo-angular-momentum-squeezed states, which we introduce in the following section.

\section{Quantum Enhancement susceptible to Loss}
\label{sec:Q-Enhancement_sus_to_loss}
In this section, we analyze the phase uncertainty of the MZI that is operated with pseudo-angular-momentum-squeezed states and driven by first-order Bragg diffraction, taking into account the loss mechanisms introduced above. We compare our results to a numerical model and provide additional analysis concerning state preparation and laser pulses.

\subsection{Phase Sensitivity for One-Axis-Twisted States}
\label{sec:OAT}
\begin{figure}[htb]
    \centering
	\includegraphics{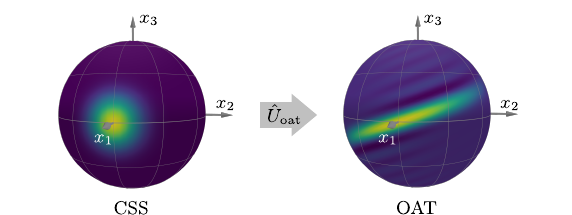}
    \caption{(Left) Angular-momentum Wigner distribution~\cite{Dowling1994-WignerDistr} of the CSS $\ket{\tfrac{\pi}{2},0}_\mt{css}$ for $N=30$ atoms [\cf{} Eq.~\eqref{eq:CSS}]. 
    It is polarized in the $x_1$-direction, with $\langle\hat{J}_1\rangle = N/2$. The angular-momentum variance is symmetrically distributed perpendicular to the $x_1$-direction, giving $\Delta J_2^2 = N/4 = \Delta J_3^2$.
    One-axis twisting creates the OAT state $\ket{\chi_{\alpha=0}}$ [\cf{} Eq.~\eqref{eq:rot_OAT} (right)]. 
    The twisting $\hat{U}_\mt{oat}$ corresponds to a $J_3$-dependent rotation distorting the CSS's symmetric distribution into an ellipse-like shape, effectively squeezing the angular-momentum variance along a direction perpendicular to $x_1$. 
    After twisting, the ellipse's major axis is inclined \wrt{} the equator. 
    Excessive twisting wraps the OAT distribution around the sphere reducing its degree of polarization. 
	}
	\label{fig:css_oat}
\end{figure}

The two-level-atom analog of the coherent states of a quantum harmonic oscillator are coherent spin states (CSSs)~\cite{Arecchi1972-CSS}.
A CSS corresponds to the $N$-atom Fock state, in which all atoms occupy the same superposition $\cos\!\frac{\Theta}{2}\hat{a}_1^\dag + \ee^{-\ii\Phi} \sin\!\frac{\Theta}{2}\hat{a}_0^\dag$ of the two momentum classes, characterized by $0\leq\Theta\leq\pi$, while $0\leq\Phi\leq 2\pi$ denotes its azimuthal polarization.
It is given by
\begin{equation}\label{eq:CSS}
	\ket{\Theta, \Phi}_\mt{css} = \frac{1}{\sqrt{N!}}\left(\cos\!\frac{\Theta}{2}\hat{a}_1^\dag + \ee^{-\ii\Phi} \sin\!\frac{\Theta}{2}\hat{a}_0^\dag\right)^N\ket{0}.
\end{equation}
Hence, the CSS is a separable state. As an example, the Fock state $\ket{N_0}$ corresponds to $\ket{0, 0}_\mt{css}$.
Figure~\ref{fig:css_oat} (left) shows the angular-momentum Wigner distribution of $\ket{\tfrac{\pi}{2},0}_\mt{css}$, which corresponds to the Fock state $\ket{N_0}$ after a beam-splitter pulse~\cite{Agarwal1981-sph_wigner_distr, Dowling1994-WignerDistr}.

The CSS does not provide quantum enhancement in the interferometer considered here as we need entanglement~\cite{Soerensen2001-Entanglement}.
Therefore, we resort to the OAT state, which is known to enable quantum enhancement of the phase sensitivity~\cite{Kitagawa1993-SSS, Pezze2018-Qrev}.
It is defined as
\begin{equation}\label{eq:rot_OAT}
	\ket{\chi_\alpha} = \ee^{-\ii \alpha \hat{S}_1}\hat{U}_\mt{oat}\ket{\tfrac{\pi}{2},0}_\mt{css},
\end{equation}
with the one-axis-twisting operator $\hat{U}_\mt{oat}(\chi) = \exp\big(-\ii\chi\hat{S}_3^2\big)$, where $\chi$ denotes the twisting strength and whose generator $\hat{S}_3^2$ is nonlinear.
The OAT transformation acts as a angular-momentum-dependent rotation that distorts the CSS Wigner distribution, squeezing the variance along one direction perpendicular to the polarization direction, while increasing it in the conjugate direction [\cf{} Fig.~\ref{fig:css_oat} (right)].
This twisting creates an ellipse-like Wigner distribution, whose major axis is inclined by an angle $\alpha_0$ (found in Appendix~\ref{app:phase_uncertainty_formulas}) \wrt{} the equator of the angular-momentum sphere. 
In comparison to the CSS's distribution, which is symmetric around the polarization direction, the OAT state's distribution is strongly dependent on its inclination \wrt{} the equator. 
This inclination angle can be controlled by a light pulse of adequate phase and pulse area~\cite{Greve2022-CavitygreveEntanglement, Corgier2021}.
The definition in Eq.~\eqref{eq:rot_OAT} therefore includes a subsequent rotation by an angle $\alpha$ about the $x_1$-axis to compensate the OAT state's initial inclination $\alpha_0$. 
According to Eq.~\eqref{eq:PS_id_MZI}, the phase uncertainty of an ideal MZI is minimized when the ellipse's major axis is aligned with the equator, \ie{} $\alpha=-\alpha_0$, as this minimizes the variance along $x_3$. 
In the following, we refer to this OAT state $\ket{\chi_{-\alpha_0}}$ as \textit{equator-OAT} state.

\subsubsection{Ideal MZI}\label{sec:ideal_MZI_PS}
The phase uncertainty of the ideal MZI operated with the equator-OAT state $\ket{\chi_{-\alpha_0}}$ is given by 
\begin{equation}
    \Delta\phi^2 = \frac{\Delta S_3^2}{\langle\hat{S}_1\rangle^2} 
    \eqcolon \frac{\xi^2}{N}
\end{equation}
where $\xi^2$ is the squeezing parameter. 
By convention, $\xi^2 < 1$ indicates that the OAT state is pseudo-angular-momentum squeezed~\cite{Wineland1992}.
Additionally, for $\xi^2 < 1$ the OAT state is entangled~\cite{Soerensen2001-Entanglement}.
For excessively large twisting strengths $\chi$, the squeezing parameter $\xi^2$ increases again because the Wigner distribution is wrapped around the sphere, reducing the projection $\langle\hat{S}_1\rangle$ and thereby degrading the phase sensitivity~\cite{Pezze2009-entanglement_and_squeezing}.
For the equator-OAT state $\ket{\chi_{-\alpha_0}}$, the angular-momentum projections and (co)variances are listed in Appendix~\ref{app:phase_uncertainty_formulas}.
The minimum of the squeezing parameter can be approximately calculated in the limit of weak twisting strengths $\chi \ll 1$ and large atom numbers $N \gg 1$, giving $\xi^2 \approx 1/2(N/3)^{-2/3}$ at a twisting strength of $\chi_0 \approx 3^{1/6}N^{-2/3}$. 
The phase uncertainty is therefore $\Delta\phi^2 = 3^{1/6}N^{-5/3}$, corresponding to quantum enhancement~\cite{Jin2009, Kitagawa1993-SSS}.
The exact value of the optimal twisting strength $\chi_0$ has to be calculated numerically.
The CSS $\ket{\tfrac{\pi}{2},0}_\mt{css}$ has a projection of $\langle\hat{S}_1\rangle = N/2$ and symmetrically distributed variances $\Delta S_2^2 = N/4 = \Delta S_3^2$, providing a phase uncertainty that saturates SNL~\cite{Pezze2018-Qrev}, \ie{} $\Delta\phi^2 = 1/N$.

\subsubsection{MZI Susceptible to Loss}\label{sec:OAT_PS}
\begin{figure}[htb]
	\centering
	\includegraphics{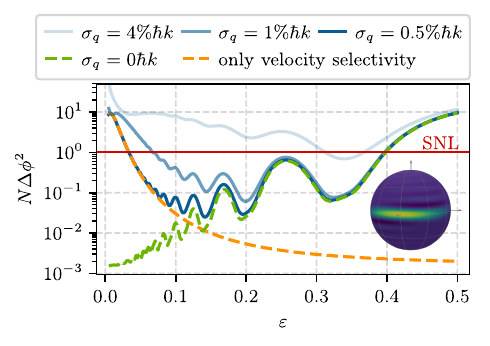}
    \caption{
        MZI phase uncertainty $N\Delta\phi^2$ normalized to SNL (red) and susceptible to velocity selectivity and parasitic diffraction to the adjacent momentum classes in dependence on the coupling strength $\varepsilon=\Omega_0/\omega_k$. 
        The sphere shows the input equator-OAT state $\ket{\chi_{-\alpha_0}}$ for $30$ atoms.
        The uncertainty is shown for numerically optimized twisting for optimal performance in an ideal MZI $\chi_0 \approx 1.61\times10^{-3}$ and $N=2\times10^4$ atoms that initially have a Gaussian momentum distribution of width $\sigma_q$ that is centered at the respective momentum class' center. 
        The uncertainty was calculated by determining the interferometric contributions from Table~\ref{tab:scr_quantities} up to the second order in $\varepsilon$.
        Dashed lines show two special cases:
        (i) No coupling to adjacent classes, only velocity selectivity for $\sigma_q=0.005\hbar k$ (orange). 
        Toward weaker coupling strengths, the system enters the velocity-selective regime, where eventually, the quantum enhancement is destroyed. 
        (ii) Including parasitic diffraction, but no velocity selectivity (green). 
        All atoms have resonant momentum $p_0$; however, the quantum enhancement is destroyed for intermediate coupling strengths in the quasi-Bragg regime.
        Competition between these two effects causes the full phase uncertainty (blue), shown for different momentum widths, to attain a minimum \wrt{} the coupling strength $\varepsilon$. 
        The phase uncertainty reaches values well below SNL; however, in both loss regimes as well as for broad momentum distributions, the quantum enhancement is destroyed.
        Due to the assumption of box-shaped pulses, the phase uncertainty oscillates with decreasing frequency for stronger $\varepsilon$.
        The lines do not show the minimal attainable phase sensitivity, as here the phases $\phi=0=\theta_0$ as well as the OAT state's inclination are fixed, although the working point is expected to drift under the influence of loss and inclination.
	}
	\label{fig:PS_Bragg_0th_2nd_order_eps}
\end{figure}

We now analyze the performance of the OAT state in the presence of velocity selectivity and parasitic diffraction.
For this analysis, we assume that the momentum distribution of the atoms is a Gaussian of standard deviation $\sigma_q$ centered at the respective center of the momentum class. The input state is taken to be the equator-OAT state $\ket{\chi_{-\alpha_0}}$ introduced above.
We set the twisting strength to the numerically obtained optimal value $\chi_0$, minimizing the squeezing parameter $\xi^2$ for best possible quantum enhancement in an ideal MZI.
The phase uncertainty of the MZI subject to atom loss is given by Eq.~\eqref{eq:PS_4x4_Bragg_MZI} and the explicit form of $\Delta\phi^2$ for the equator-OAT state is given in Eq.~\eqref{aeq:ps_alpha_equator_OAT}.
The results are shown in Fig.~\ref{fig:PS_Bragg_0th_2nd_order_eps}.
The dashed lines represent the limits of negligible parasitic diffraction (orange, velocity selectivity only) and negligible velocity selectivity (green, parasitic diffraction only).
These lines illustrate the two loss regimes: (i) The velocity-selective regime for weak coupling strengths $\varepsilon$. 
Here, atoms are lost to the quasi-incoherent background and the phase sensitivity is further degraded by a population imbalance on the interferometer arms.
(ii) The quasi-Bragg regime~\cite{Mueller2008-adiab_elim_jBragg}, where atom loss to the adjacent momentum classes dominates velocity selectivity for stronger coupling strengths $\varepsilon$.
The competition between these two effects gives rise to the full phase uncertainty (solid blue lines), shown for different momentum widths $\sigma_q$.
The red line denotes SNL. 
We find that the equator-OAT state still enables quantum enhancement, which is, however, destroyed in both loss regimes, such that only intermediate coupling strengths allow for a phase uncertainty below SNL.
Moreover, the momentum distribution must be sufficiently narrow, as the overall phase sensitivity quickly deteriorates for larger $\sigma_q$, underlining the need for advanced cooling techniques such as delta-kick collimation~\cite{Muentinga2013-Bragg_AI,Kovachy2015_Matter_wave_lensing}.
Finally, the oscillatory dependence of the phase uncertainty on the Rabi frequency arises from the box-shaped light pulses.
We generalize this analysis by numerically implementing time-dependent pulses below.
Since the optimal phase sensitivity occurs at locally pronounced minima in dependence on $\varepsilon$, it becomes strongly sensitive to laser-intensity fluctuations.

\subsubsection{Comparison to Numerical Model}
\label{sec:num_sol}
\begin{figure}[htb]
	\centering
	\includegraphics{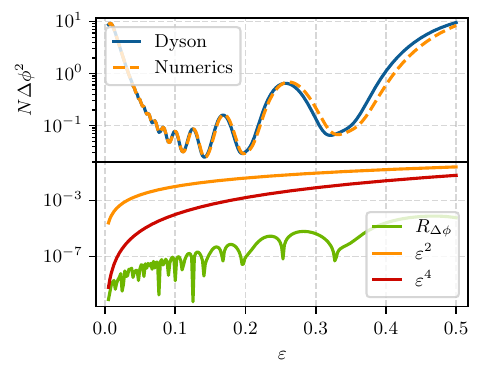}
	\caption{
        (Top panel) Phase uncertainty of the MZI evaluated at phases $\phi=0=\theta_0$ and seeded by the equator-OAT state $\ket{\chi_{-\alpha_0}}$, twisted to perform optimally in an ideal MZI ($\chi_0\approx1.61\times10^{-3}$), for $N=2\times10^4$ atoms with a Gaussian momentum distribution of width $\sigma_q=0.005\hbar k$ centered in the respective momentum class.
        The perturbative calculation $\Delta\phi^2$ (blue) shows excellent agreement with the numerical calculation for the six-momentum-class system $\Delta\phi^2_\mt{num}$ (dashed orange).
        (Bottom panel) Residual $R_{\Delta\phi}= |\Delta\phi^2 - \Delta\phi^2_\mt{num}|$ of the perturbative calculation. 
        Functions $\varepsilon^2$ and $\varepsilon^4$ confirm the validity of the perturbative calculation up to fourth order in the coupling strength $\varepsilon$.
        This high accuracy is provided by the Eq.~\eqref{eq:PS_4x4_Bragg_MZI} effectively corresponding to a Pad\'{e} approximation.
	}
	\label{fig:PS_Bragg_num_vs_ana}
\end{figure}

To validate the perturbative results shown in Fig.~\ref{fig:PS_Bragg_0th_2nd_order_eps}, we compare them with a numerical model.
We consider resonant first-order Bragg diffraction and truncate the system to six momentum classes $n=-2,-1,0,1,2,3$. 
This extended truncation verifies that the four-class description used in the perturbative treatment is sufficient to obtain an accurate description of the main momentum classes $n=0,1$ up to second order in the coupling strength $\varepsilon$.
The dynamics of the field operators $\hat{\Psi}_n$ are governed by the equation of motion given by Eq.~\eqref{eq:Heisenberg_eom}, where now the summation extends over all six momentum classes.
We use the standard Python Runge--Kutta solver \texttt{solve\_ivp} from the \texttt{SciPy} package~\cite{Scipy} to solve the equations of motion and obtain the reflection and transmission coefficients of the beam-splitter and mirror pulses entering the MZI time evolution in analogy to Eq.~\eqref{eq:Heisenberg_pic_field_ops}.
The MZI is operated with the equator-OAT state, which is twisted to perform optimal in the ideal MZI, and we employ the same detection scheme as above, measuring $\hat{J}_{3,\mt{H}}$ while excluding the outer exit ports.
Figure~\ref{fig:PS_Bragg_num_vs_ana} (top panel) compares the numerical result for the phase uncertainty to the perturbative expression and shows the residual $R_{\Delta\phi}= |\Delta\phi^2 - \Delta\phi^2_\mt{num}|$ (bottom panel). 
As a reference for the validity of the perturbative calculation, the latter also includes $\varepsilon^2$ and $\varepsilon^4$.
According to the residual, the perturbative calculation is exact even to fourth order in $\varepsilon$.
This accuracy seems to be consistent with the Pad\'{e}-like structure of the phase uncertainty from Eq.~\eqref{eq:PS_4x4_Bragg_MZI}, where the script quantities are determined up to $\mathcal{O}(\varepsilon^3)$ without secular terms in our perturbative treatment.
Consequently, the perturbative calculation agrees with the numerical model surprisingly well even for intermediate coupling strengths $\varepsilon$ in the quasi-Bragg regime, despite the atom-light interaction being calculated only up to second order Dyson.

\subsection{Blackman Pulses}
\label{sec:Blackman-pulses}
The oscillatory dependence of the phase sensitivity on $\varepsilon$ in Fig.~\ref{fig:PS_Bragg_0th_2nd_order_eps} arises from the assumption of box-shaped pulses.
To assess the role of the pulse shape, we use the numerical model form Sec.~\ref{sec:num_sol} and implement a time-dependent Rabi frequency $\varepsilon_\mt{td}(\lambda) = \Omega_0(\lambda)/\omega_k$ in form of a Blackman pulse, which is given by
\begin{equation}
	\varepsilon_\mt{td}(\lambda) = \varepsilon 
	\left( 
	0.42 
	- 0.5 \cos \frac{2 \pi \lambda}{\lambda_j}
	+ 0.08 \cos \frac{4 \pi \lambda}{\lambda_j}
	\right),
\end{equation}
for $0\leq\lambda\leq\lambda_j$ and otherwise $\varepsilon_\mt{td}(\lambda)=0$, where $\lambda_j$ is the dimensionless pulse duration.
The maximal coupling strength is denoted by $\varepsilon$.
The pulse length satisfies $\lambda_j = \tau/(a_0\varepsilon)$, with pulse area $\tau = \int\!\dd \lambda\, \varepsilon_\mt{td}(\lambda)$.
Again, we assume that the beam-splitter and mirror pulse have the same maximal coupling strength, where the latter is realized by doubling the pulse duration.
\begin{figure}[b]
	\centering
	\includegraphics{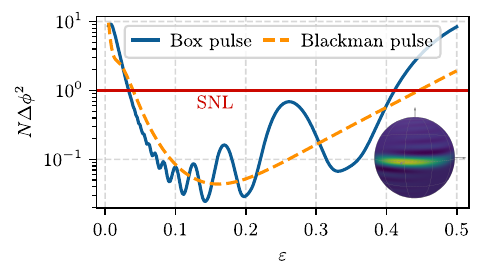}
	\caption{
        Phase uncertainty obtained when using box-shaped (blue) and Blackman pulses (dashed orange) in an MZI seeded by the equator-OAT state $\ket{\chi_{-\alpha_0}}$, twisted to perform optimally in an ideal MZI ($\chi_0\approx1.61\times10^{-3}$), at $\phi=0=\theta_0$ for $N=2\times10^4$ atoms with a Gaussian momentum distribution centered at each momentum class center, with standard deviation $\sigma_q=0.005\hbar k$.
    	The numerically obtained phase uncertainty using Blackman pulses does not exhibit oscillations with $\varepsilon$ and shows a clear minimum, rendering it more robust to laser-intensity fluctuations. 
        However, Blackman pulses perform slightly worse in the velocity-selective regime, while being partially better in the quasi-Bragg regime.
    }
	\label{fig:PS_Bragg_box_vs_blackman}
\end{figure}
Figure~\ref{fig:PS_Bragg_box_vs_blackman} displays the phase uncertainty of an MZI operated with the equator-OAT state for both box-shaped and Blackman pulses, revealing that the uncertainty does not oscillate with the coupling strength for Blackman-shaped pulses.
The Blackman pulse's performance in the velocity-selective regime is slightly inferior to that of the box-shaped pulse.
This effect is caused by their momentum-dependent reflectivities $|R_j|^2$ and $|\tilde{R}_j|^2$ being narrower than for box-shaped pulses, resulting in stronger velocity selectivity.
In the quasi-Bragg regime, the uncertainty follows the same trend as for box-shaped pulses, but without the pronounced oscillations.
Accordingly, the phase uncertainty exhibits a well-defined minimum associated with a unique optimal coupling strength, rendering it less sensitive to laser-intensity fluctuations or beam profiles.
The lower phase uncertainty for larger interaction strengths can be explained by suppressed higher-order diffraction, due to the smooth ramp-up of the light field. 
Overall, Blackman pulses provide quantum enhancement in the same order of magnitude as box-shaped pulses.

\subsection{Comparison to CSS}
\label{sec:Comp-to-CSS}
In an ideal MZI, the CSS provides shot-noise-limited sensitivity.
To compare its performance under atom loss with that of the equator-OAT state, whose inclination is fixed at $\varphi=0$, we consider the CSS $\ket{\frac{\pi}{2},0}_\mt{css}$ at the same operating point $\phi = 0 = \theta_0$. We assume that the momentum distribution of the atoms is a Gaussian, with standard deviation $\sigma_q$, that is centered in each momentum class. 
The corresponding phase uncertainty is obtained by taking Eq.~\eqref{aeq:ps_alpha_equator_OAT} and setting the twisting strength to $\chi = 0$.
\begin{figure}[b]
	\centering
	\includegraphics{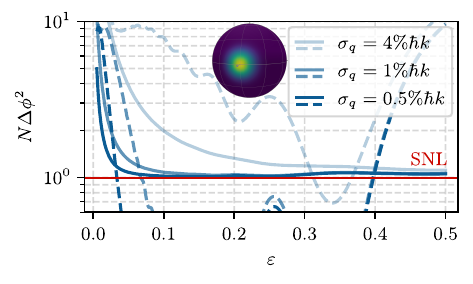}
    \caption{
        Phase uncertainty of the MZI evaluated at phases $\phi=0=\theta_0$ with fixed inclination $\varphi = 0$ and driven by the CSS $\ket{\frac{\pi}{2},0}_\mt{css}$ (displayed on sphere $N=30$) for $N=2\times10^4$ atoms with a Gaussian momentum distribution centered in the respective momentum classes, with different standard deviations $\sigma_q$ (solid blue). 
        Here, $\varepsilon = \Omega_0/\omega_k$ is the coupling strength.
        In the velocity-selective regime, the CSS's phase sensitivity deteriorates quickly for weaker coupling strengths, whereas in the quasi-Bragg regime it is only mildly affected.
        Overall, the CSS does not exhibit $\varepsilon$-dependent oscillations, thereby being robust against laser-intensity fluctuations. 
        Additionally, the equator-OAT state $\ket{\chi_{-\alpha_0}}$ (dashed blue), twisted to perform optimally in an ideal MZI ($\chi_0\approx1.61\times10^{-3}$), performs worse than the CSS in both loss regimes at $\phi=0=\theta_0$ and zero inclination. 
	}
	\label{fig:PS_fock_state}
\end{figure}
Figure~\ref{fig:PS_fock_state} shows the phase uncertainty (solid lines) together with the equator-OAT results (dashed lines) for reference.
We emphasize that the comparison presented is performed at the fixed operating point $\phi = 0 = \theta_0$ with fixed inclination $\varphi = 0$.
Under these operating conditions, the CSS outperforms the equator-OAT state.
However, optimizing the inclination $\varphi$ allows the OAT state to surpass the CSS in both regimes, as discussed in Sec.~\ref{sec:inclination-optimization}.
While the choice $\phi = 0$ minimizes the phase uncertainty in the ideal MZI, the presence of loss shifts this working point, degrading the OAT state's phase sensitivity.
This drift of the working point can be partially mitigated by optimizing the initial inclination of the OAT state \wrt{} the angular-momentum sphere's equator, as we detail in Sec.~\ref{sec:inclination-optimization}.
Figure~\ref{fig:PS_fock_state} furthermore shows that for sufficiently narrow momentum distributions, the CSS saturates the SNL over a broad range of coupling strengths and is less sensitive to laser-intensity fluctuations compared to the OAT.

\subsection{Input-State Optimization}
\label{sec:inclination-optimization}
For the ideal MZI, we found that the OAT state performs best if it is oriented along the equator of the angular-momentum sphere ($\alpha = -\alpha_0$) and twisted such that the squeezing parameter $\xi$ becomes minimal ($\chi = \chi_0$).
In Sec.~\ref{sec:OAT_PS}, we adopted this configuration to compute the phase uncertainty under the influence of atom loss.
However, the beam-splitter and mirror transformations are strongly dependent on velocity selectivity and parasitic diffraction, suggesting this choice to be suboptimal once loss mechanisms are included.
To investigate this aspect, we numerically optimize the phase uncertainty given in Eq.~\eqref{aeq:ps_alpha_equator_OAT} \wrt{} to the inclination angle $\varphi = \alpha + \alpha_0$ of the OAT state's major axis relative to the equator of the sphere and also separately \wrt{} the twisting strength.

\begin{figure}[htb]
	\centering
	\includegraphics{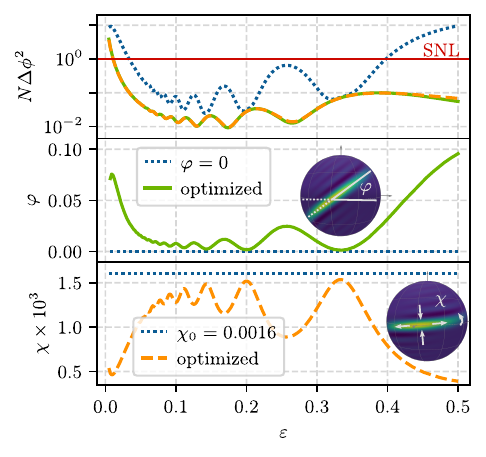}
    \caption{
        (Top panel) Phase uncertainty of the MZI evaluated at phases $\phi=0=\theta_0$ for $N=2\times10^4$ atoms with a Gaussian momentum distribution centered at each momentum class center, with standard deviation $\sigma_q=0.005\hbar k$. 
        The MZI is driven by the equator-OAT state $\ket{\chi_{-\alpha_0}}$, that is twisted to perform optimal in the ideal MZI, \ie{} $\chi_0\approx1.61\times10^{-3}$ (blue), inclination-optimized-OAT state (green) and the twisting-optimized-OAT state (orange). 
        For the latter, we fix the initial rotation $\alpha$, such that for optimal twisting $\chi_0$ in the ideal MZI, the OAT ellipse's major axis is aligned with the equator and we then optimize \wrt{} twisting strength.
        The former minimizes the phase uncertainty Eq.~\eqref{aeq:ps_alpha_equator_OAT} \wrt{} the inclination $\varphi = \alpha + \alpha_0$, \cf{} Appendix~\ref{app:phase_uncertainty_formulas}. 
        (center panel) Corresponding optimal inclination $\varphi$.
        Optimization of the OAT state's initial inclination substantially enhances the phase sensitivity compared to equator alignment $\varphi=0$. 
        This is especially effective in the quasi-Bragg regime, whereas in the velocity-selective regime, the momentum-dependent beam-splitter and mirror pulses smear out the OAT Wigner distribution, rendering the optimization less effective.
        (bottom panel) Optimal twisting strength $\chi$. Optimizing \wrt{} twisting strength effectively sets the optimal inclination, while sacrificing degree of squeezing, leading to the twisting optimization to be less efficient than the inclination optimization.
    }
	\label{fig:ps_orientation_opt}
\end{figure}

First, we optimize \wrt{} the inclination, while fixing the twisting to the optimal value $\chi_0$ in the ideal MZI.
We again assume a Gaussian momentum distribution centered in the respective momentum class with a standard deviation of $\sigma_q$.
The results are shown in Fig.~\ref{fig:ps_orientation_opt} (top panel), where we display the equator-OAT state (dotted blue) and the inclination-optimized-OAT state (green) as well as the corresponding inclination $\varphi$ (central panel).
The inclination optimization substantially reduces the uncertainty, especially in the quasi-Bragg regime.
Atom loss during the interferometric sequence modifies the squeezing properties of the OAT state, particularly its inclination against the equator, thereby effectively shifting the working point from the optimal choice $\phi=0$ in the ideal MZI and degrading phase sensitivity. 
This drift can be partially compensated by adjusting the initial inclination of the OAT state~\cite{Goldsmith2025-Derckeff, Greve2022-CavitygreveEntanglement, deAraujo2026}.
In the velocity-selective regime, momentum-dependent beam-splitter and mirror transformations smear out the OAT state's Wigner distribution, reducing the degree of squeezing and thus the phase uncertainty's sensitivity to the initial inclination. 
This effect is absent in the quasi-Bragg regime ($\varepsilon\lesssim1$), where atom loss is suspected to influence the inclination of the OAT state during the interferometer sequence, strongly affecting the phase uncertainty.
Consequently, as shown in Fig.~\ref{fig:ps_orientation_opt}, the inclination optimization substantially enhances the sensitivity in the quasi-Bragg regime.
Compared to the CSS's performance (\cf{} Fig.~\ref{fig:PS_fock_state}), the inclination-optimized-OAT state performs better across all coupling strengths considered and allows for operation below shot noise over a large parameter range.
The center panel in Fig.~\ref{fig:ps_orientation_opt} reveals that already minute inclination changes have a strong effect on the phase uncertainty. 
This stems from the OAT state's highly elliptical shape, whose variance along $x_3$ increases rapidly under a slight inclination \wrt{} the equator.

Alternatively, the phase uncertainty can be optimized \wrt{} the twisting strength $\chi$.
Here, we fix the initial rotation $\alpha$ of the OAT state such that using optimal twisting strength $\chi_0$ in the ideal MZI yields inclination $\varphi = 0$. 
The phase uncertainty Eq.~\eqref{aeq:ps_alpha_equator_OAT} is then minimized \wrt{} the twisting strength $\chi$.
The result is depicted in Fig.~\ref{fig:ps_orientation_opt} (top panel, dashed orange), with corresponding optimal twisting strength (bottom panel).
We observe an increase in sensitivity, especially in the quasi-Bragg regime, which has been observed before \cite{Günther2024-squeezingenhancementlossymultipath}. 
Generally, the optimization's behavior is extremely similar to that of the inclination optimization, but the former performs marginally worse.
The phase uncertainty is strongly sensitive to small changes in the inclination of the OAT state \wrt{} the equator, and rather robust to small changes in the OAT ellipse's shape set by the twisting strength~\cite{deAraujo2026}.
Furthermore, the initial inclination $\alpha_0$, included in the OAT-transformation $\hat{U}_\mt{oat}$, is strongly dependent on the twisting strength.
Therefore, setting a twisting slightly different from $\chi_0$ marginally affects the OAT state's shape, implying, however, a change in inclination comparable to the optimal value shown in Fig.~\ref{fig:ps_orientation_opt} (center panel).
Consequently, twisting optimization effectively reduces to inclination optimization.
This connection is confirmed by the strong correlation between the $\varepsilon$ dependence of the optimal inclination and twisting strength.
As shown in Fig.~\ref{fig:ps_orientation_opt} (bottom panel), this optimal inclination is reached for twisting strengths $\chi<\chi_0$, \ie{} by sacrificing degree of squeezing, which accounts for the slight performance difference relative to inclination optimization.
In an experimental implementation, the twisting strength should therefore be tuned to the optimal value for an ideal MZI and the pulse duration of the first beam-splitter pulse needs to be tuned to realize the optimal inclination $\varphi$ shown in Fig.~\ref{fig:ps_orientation_opt} (center panel)~\cite{Greve2022-CavitygreveEntanglement}.

In principle, simultaneously optimizing the interferometer phase and the first beam-splitter phase, which we set to $\phi = 0 = \theta_0$ for simplicity, could further enhance the phase sensitivity; however, we suspect the attainable enhancement to be comparable to that of the inclination optimization.

\begin{figure}[b]
    \centering
    \includegraphics{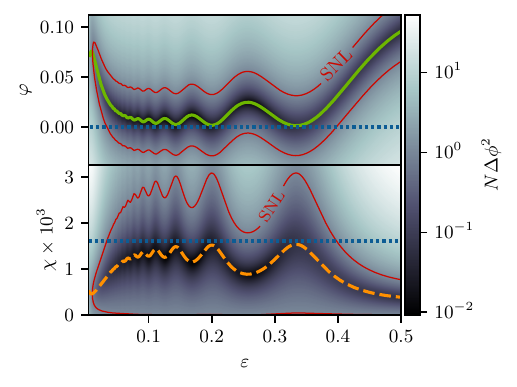}
    \caption{
        Quantum-enhanced phase uncertainty of the MZI at phases $\phi = 0 = \theta_0$, atom number $N=2\times10^4$ and a momentum standard deviation of $\sigma_q = 0.005\hbar k$ for different values of the coupling strength $\varepsilon$. 
        The red line shows the SNL, \ie{} $N\Delta\phi^2 = 1$.
        (Top panel) Dependence on the inclination $\varphi = \alpha + \alpha_0$ with the twisting strength set to the optimal value in an ideal MZI, \ie{} $\chi_0\approx1.61\times10^{-3}$. 
        (Green) Minimum of the phase uncertainty \wrt{} inclination. 
        (Dotted blue) Choice of optimal inclination in the ideal MZI.
        (Bottom panel) Dependence on the twisting strength $\chi$, where the initial rotation $\alpha$ is chosen such that for optimal twisting $\chi_0$ in the ideal MZI, the inclination is $\varphi=0$. 
        (Dashed orange) Minimum of the phase uncertainty \wrt{} twisting strength.
        (Dotted blue) Choice of optimal twisting strength in the ideal MZI. 
        }
    \label{fig:full_optimized_PU_twisting_inclionation}
\end{figure}
To study the susceptibility to inclination and twisting-strength fluctuations, we show the phase uncertainty in two density plots in Fig.~\ref{fig:full_optimized_PU_twisting_inclionation} at the same working point as in Fig.~\ref{fig:ps_orientation_opt}.
The top panel depicts different inclinations $\varphi$ while the bottom panel highlights the susceptibility \wrt{} the twisting strength $ \chi$.
The solid green and dashed orange lines indicate the optimal inclination and twisting strength, respectively, while the red lines mark the boundary of the SNL. 
We find that, over most of the investigated range of $\varepsilon$, maintaining sub-shot-noise sensitivity requires the inclination to remain within approximately 0.025\,rad of its optimal value. 
Similarly, fluctuations of the twisting strength exceeding the order of $10^{-3}$ can lead to a loss of quantum enhancement.

\section{Discussion}
\label{sec:Discussion}
This study presents an analytical expression for the phase uncertainty of a first-order Bragg MZI. 
Our formalism fully accounts for velocity selectivity and incorporates diffraction to higher orders via perturbation theory, assuming box-shaped light pulses.
We have demonstrated that spatially cropping the detection zone enhances the sensitivity in the velocity-selective regime. 
Furthermore, including atom loss reveals that quantum enhancement from OAT states can only be harnessed in an intermediate regime of the pulse duration, balancing parasitic diffraction and velocity selectivity.
The enhancement can only be provided by sufficiently cold atomic ensembles, emphasizing the need for cooling methods like delta-kick collimation~\cite{Muentinga2013-Bragg_AI,Kovachy2015_Matter_wave_lensing}.
Finally, input-state optimization was identified as a crucial requirement for sensitivity enhancement in experimental implementations.

Our detection scheme in the velocity-selective regime requires detection after an intermediate time of flight to spatially resolve and distinguish the momentum classes. 
Cropping the outer exits can only be done if the distinct atomic clouds do not overlap. 
If overlap occurs, it introduces a beating into the signal significantly deteriorating sensitivity \cite{Jenewein2022-Spurious_paths_Bragg_AI,Kirsten-Siemss2023}.
In an MZI subject to velocity selectivity and parasitic diffraction, the CSS performs shot-noise limited over a large range of coupling strengths, while being robust against laser-intensity fluctuations.
Conversely, while the OAT state potentially performs better than the CSS, the associated uncertainty is highly sensitive to its inclination~\cite{Goldsmith2025-Derckeff, Greve2022-CavitygreveEntanglement}.
Therefore, the OAT state requires careful and low-noise state preparation with optimal inclination. 
However, atom-number fluctuations and deviations in the twisting strength induce inclination fluctuations limiting the minimal attainable uncertainty. 
Similarly, fluctuations in the laser intensity influence the pulse area controlling the inclination.
Furthermore, we assume that atom-atom interactions during the interferometer sequence are negligible.
This requires that the density has decreased sufficiently after the state-preparation stage such that the residual collisional nonlinearities do not induce additional one-axis twisting during the interferometer.
Otherwise, the squeezing dynamics continues during the sequence and the optimal inclination must be reassessed to account for the modified nonlinear evolution~\cite{Feldmann2023-optimalsqueezinghighprecisionatom}, making the inclination optimization dependent on the residual interaction dynamics.
In practice, the squeezing preparation stage may require an additional evolution time before the sensing sequence begins. 
For example, in a preceding state-engineering interferometer, the preparation of the desired squeezing for \rb{} requires an additional timescale on the order of tens of milliseconds~\cite{Szigeti2020-OAT_MZI}, after which the atomic density has decreased sufficiently that interactions become negligible.
This contribution increases the total sequence duration.
Conversely, the required preparation time may be reduced by increasing the interaction strength and density for a short period of time during the squeezing stage through techniques such as delta-kick squeezing~\cite{Corgier2021}.

While inclination optimization has been experimentally demonstrated in a cavity-mediated OAT setup using Raman diffraction \cite{Greve2022-CavitygreveEntanglement}, these systems introduce unique challenges. 
Specifically, cavity-converted frequency noise drives parasitic Bragg diffraction, resulting in atom loss to higher orders and thereby signal deterioration. 
The analytical framework presented here can be generalized to describe similar situations, effectively quantifying atom loss (and phase shifts).

Like previous approaches~\cite{Günther2024-squeezingenhancementlossymultipath} based on the Landau--Zener formalism~\cite{Siemss2020-JulianBackgrouondPhysRevA.102.033709}, our treatment can be generalized to higher-order Bragg diffraction.
While extending the perturbative formalism~\cite{SIso2018} to $n$th-order diffraction is in principle possible, higher diffraction orders introduce additional spurious paths and associated beatings. 
These effects require in general a multi-mode treatment beyond the present framework and can degrade the phase sensitivity.
Consequently, next steps might involve incorporating mitigation methods, such as dichroic mirror pulses~\cite{Pfeiffer2025-DominikDchrMirr, Siemss2020-JulianBackgrouondPhysRevA.102.033709} or optimal control techniques~\cite{Wang2024, Saywell2023, Fang2018-OTC, Wilkason2022, Rodzinka2024}, to suppress population of spurious paths or their deleterious effects.
Such scenarios require a more careful second-quantized analysis, as cropping deteriorating contributions from the signal becomes more complex.
Inherently, such a formalism has to be based on a multi-mode treatment to incorporate loss modes, similar to the four-mode system described here.
Analogous treatments are necessary to describe large-momentum transfer facilitated by Bloch oscillations, which is also an inherently multi-mode system~\cite{Fitzek2024-BO_LMT}.  
Finally, our numerical results demonstrate the importance of extending this analysis to time-dependent pulses. 
An analytical treatment of time-dependent pulses would, however, require incorporating Doppler detuning, \ie{} velocity selectivity, as a perturbation.
Similarly, our treatment can in principle be generalized to pulse sequences beyond an MZI and including the squeezing within the sequence~\cite{Linnemann2016_Quantum-Enhanced} or to echo sequences targeting dephasing noise~\cite{Schulte2020, Bringewatt2026}. 
Other natural extensions include Ramsey-Bord\'e-type geometries~\cite{Borde1989}, where atoms are inherently lost during the sequence by design. 
These are often combined with Bloch oscillations for precision measurements of the fine-structure constant, which continue to face systematic discrepancies~\cite{Parker2018-FineStr_const,Morel2020}. 

Here, we focused on the OAT state to generate quantum enhancement in the MZI, but in principle also twin-Fock states are a viable option~\cite{Lücke2011-Q_enh_Twin_Fock, Lange2018-Q-Enh}.
However, as our analysis relies on the first moment of the pseudo-angular-momentum operator, twin-Fock states, despite their strong entanglement, remain insensitive to the interferometric phase.
Their quantum advantage emerges only at the level of higher-order moments~\cite{Kim1998-NoNeqN}, which are straightforwardly included in our formalism.

\begin{acknowledgements}
We thank J. G\"unther, N. Gaaloul, and K. Hammerer, as well as F. Schmidt-Kaler and C. Klempt for helpful feedback.
We are also grateful to the QUANTUS and INTENTAS teams for helpful discussions.
The INTENTAS and QUANTUS projects are supported by the German Aerospace Center [Deutsches Zentrum für Luft- und Raumfahrt (DLR)] with funds provided by the Federal Ministry of Economic Affairs and Energy [Bundesministerium für Wirtschaft und Energie (BMWi)] and the Federal Ministry for Economic Affairs and Climate Action [Bundesministerium für Wirtschaft und Klimaschutz (BMWK)] due to an enactment of the German Bundestag under grant nos. 50WM2177 (INTENTAS) and 50WM2450E (QUANTUS-VI). 
\end{acknowledgements}

\section*{Author Contributions}
Conceptualization: CK (supporting), DD, EG;
Data curation: CK; 
Formal analysis: CK;
Funding acquisition: EG;
Methodology: CK (lead), DD, EG;
Supervision: DD, EG;
Validation: CK, DD, EG;
Visualization: CK (lead), DD, EG;
Writing -- original draft: CK;
Writing -- review \& editing: CK, DD, EG.

\section*{Data Availability}
All data that support the findings are openly available~\cite{Karres2026}.

\appendix

\section{Secular-term-free Perturbative Description of first-order Bragg Diffraction}
\label{app:Pert-Bragg}

\subsection{Dyson in Suitable Picture}
We consider resonant first-order Bragg diffraction between four momentum classes $n=-1,0,1,2$, which is described by the Hamiltonian given in Eq.~\eqref{eq:Bragg_final_Hamiltonian}
\begin{equation}
    \hat{H} = \hbar \omega_k\sum_{m,n}\int_{I_k}\!\dd q\, \hat{\Psi}_m^\dag \mathcal{H}_{mn} \hat{\Psi}_n^{\phantom{\dag}}.
\end{equation}
We regard the coupling to the adjacent classes $n=-1,2$ as a perturbation and write the dimensionless Hamiltonian as
\begin{equation}
	\mathcal{H} = \mathcal{H}_0 + \varepsilon \mathcal{V},
\end{equation}
where
\begin{equation}
    \mathcal{H}_0 = \frac{1}{2}
    \begin{pmatrix}
		2\delta_{-1} & 0 & 0 & 0 \\
		0 & 2\delta_0 & \varepsilon\ee^{-\ii\theta} & 0 \\
		0 & \varepsilon\ee^{\ii\theta} & 2\delta_1 & 0 \\
		0 & 0 & 0 & 2\delta_{2}
	\end{pmatrix},
\end{equation}
describes the unperturbed system, \ie{} the coupling between the main momentum classes $n=0,1$ and the detunings to the adjacent classes.
The perturbation $\mathcal{V}$ is given by
\begin{equation}
    \mathcal{V} = \frac{1}{2}
    \begin{pmatrix}
		0 & \ee^{-\ii\theta} & 0 & 0 \\
		\ee^{\ii\theta} & 0 & 0 & 0 \\
		0 & 0 & 0 & \ee^{-\ii\theta} \\
		0 & 0 & \ee^{\ii\theta} &0
	\end{pmatrix},
\end{equation}
and the coupling strength $\varepsilon = \Omega_0/\omega_k$ corresponds to the dimensionless Rabi frequency.
The detuning $\delta_n = \left(n-1/2\right) \varepsilon v + n (n-1)$ depends on the dimensionless Doppler detuning $v(q) = \nu_k(q)/\Omega_0$ and, notably, on $\varepsilon$.
Given that $\varepsilon$ will serve as the expansion parameter, we require the condition $\varepsilon\ll1$.
Consequently, this definition necessitates $\varepsilon v = \nu_k/\omega_k \ll 1$. 
This formulation of $\delta_n$ substantially simplifies the expansion of the couplings in $\varepsilon$, while the required condition is typically fulfilled in standard application scenarios.

Following standard procedure and calculating the time evolution using the Dyson series in the interaction picture \wrt{} $\mathcal{H}_0$ results in secular terms $\propto \lambda^n\varepsilon^m$, proportional to powers of the dimensionless time $\lambda = \omega_k t$. 
For an accurate description of this system, these terms can be systematically eliminated by transforming into a suitable interaction picture~\cite{SIso2018}.
We call the generator of this image transformation $\tilde{\mathcal{H}}_0$ and rewrite the Hamiltonian accordingly
\begin{equation}
	\mathcal{H} = \tilde{\mathcal{H}}_0(\varepsilon) + \varepsilon \tilde{\mathcal{V}} (\varepsilon),
\end{equation}
where $\tilde{\mathcal{V}}$ is the adapted perturbation.
We aim to describe the coupling up to $\mathcal{O}(\varepsilon^3)$; hence, we have to eliminate secular terms up to that same magnitude. 
In our case, the generator and adapted perturbation are
\begin{equation}
	\tilde{\mathcal{H}}_0(\varepsilon) = \mathcal{H}_0 + \varepsilon^2 \mathcal{H}_2(\varepsilon)\quad\text{and}\quad \tilde{\mathcal{V}}(\varepsilon) =  \mathcal{V} - \varepsilon \mathcal{H}_2(\varepsilon),
\end{equation}
where $\mathcal{H}_2$ is tailored to eliminate secular terms of relevant order and takes the form
\begin{equation}\label{aeq:H_2}
	\mathcal{H}_2(\varepsilon) = \frac{1}{16}
    \begin{pmatrix}
		2 + \varepsilon v & 0 & 0 & 0 \\
		0 & -\varepsilon v\beta - 2 & \varepsilon\ee^{-\ii\theta}\beta & 0 \\
		0 & \varepsilon\ee^{\ii\theta}\beta & \varepsilon v\beta - 2 & 0 \\
		0 & 0 & 0 & 2 - \varepsilon v
	\end{pmatrix}.
\end{equation}
Here, $\beta(q) = (v/f)^2 - 1/(2f^2)$ depends on the dimensionless effective Rabi frequency $f(q)=\sqrt{1 + v(q)^2}$ and the dimensionless Doppler detuning $v(q) = \nu_k(q)/\Omega_0$.
The matrix elements of $\mathcal{H}_2$ reflect the secular terms we aim to eliminate.
Equation~\eqref{aeq:H_2} shows that $\mathcal{H}_2$ has a linear dependence on $\varepsilon$.
This is due to the fact that we also need to eliminate secular terms $\propto\lambda\varepsilon^3$, as the pulse area is given by $\tau = \lambda \varepsilon$, which corresponds to $\mathcal{O}(\varepsilon^0)$. 

Transforming into the interaction picture via the unitary $\tilde{\mathcal{U}}_0 = \exp\bigl(-\ii \tilde{\mathcal{H}}_0\lambda\bigr)$ gives the corresponding Hamiltonian
\begin{equation}
	\mathcal{H}_\mt{I} = \varepsilon\mathcal{V}_\mt{I} - \mathcal{H}_2,
\end{equation}
where we have used $[\mathcal{H}_0,\mathcal{H}_2] = 0$ and defined the transformed perturbation as $\mathcal{V}_\mt{I} = \tilde{\mathcal{U}}_0^\dag \mathcal{V} \tilde{\mathcal{U}}_0^{\phantom{\dag}}$.
The time-evolution operator in dimensionless time is then given by 
\begin{equation}
	\mathcal{U}(\lambda) = \mathcal{T}\exp\left\{ -\ii \int\limits_{0}^{\lambda}\!\dd s\, \mathcal{H}_\mt{I}(s) \right\},
\end{equation}
and can be approximated by calculating the Dyson series up to second order, which gives
\begin{align}\label{aeq:U_Dyson}
\begin{split}
	\mathcal{U}(\lambda) \approx &\id - \ii\varepsilon \int\limits_{0}^{\lambda} \! \dd \lambda_1\, \mathcal{V}_{\mathrm{I}}(\lambda_1) 
	+ \ii\varepsilon^2 \int\limits_{0}^{\lambda} \! \dd \lambda_1\, \mathcal{H}_2
    \\
    &
	\quad + \left( -\ii \varepsilon \right)^2 \int\limits_{0}^{\lambda} \! \dd \lambda_1 \int\limits_{0}^{\lambda_1} \! \dd \lambda_2\, \mathcal{V}_{\mathrm{I}}(\lambda_1) \mathcal{V}_{\mathrm{I}}(\lambda_2).
\end{split}
\end{align}
Here, the second term eliminates secular terms that are created in the third. Terms one would usually expect to appear at second-order Dyson, \ie{} $\varepsilon^4\lambda^2\mathcal{H}_2^2$, $\varepsilon^3\iint\mathcal{H}_2\mathcal{V}_\mt{I}$ and $\varepsilon^3\iint\mathcal{V}_\mt{I}\mathcal{H}_2$, are either of higher order in $\varepsilon$ or canceled in higher orders of the Dyson series~\cite{SIso2018}.

The couplings calculated via Eq.~\eqref{aeq:U_Dyson} consist of $\varepsilon$-dependent amplitudes and trigonometric functions with $\varepsilon$-dependent frequencies.
To obtain a secular-term-free description accurate to $\mathcal{O}(\varepsilon^3)$, the amplitudes are expanded in powers of $\varepsilon$, while the trigonometric functions are preserved, as their expansion reintroduces secular terms.
The latter is especially relevant when transforming back into the initial picture, in which the system is described by $\mathcal{H}$.

\subsection{Solution for first-order Bragg Diffraction}
\label{asec:pert_sol_bragg_diff}
Here, we list the solution for resonant first-order Bragg diffraction. 
In the Heisenberg picture, the time evolution of the main momentum classes $n=0,1$ takes the form 
\begin{subequations}\label{aeq:Psi_main_Heisenberg}
	\begin{equation}
		\hat{\Psi}_{m,\mt{H}}(q,\lambda)\ket{\psi_\mt{in}} = 
		  \sum_{n=0}^1 G_{mn} \hat{\Psi}_n(q,0)\ket{\psi_\mt{in}},
	\end{equation}
    where $m = 0,1$ and
	\begin{equation}
	G(q,\tau,\theta) = 
    \begin{pmatrix}
        T & R \\
        \tilde{R} & \tilde{T} \\
    \end{pmatrix} =
    \begin{pmatrix}
        \tilde{t} & \tilde{r} \\
        -\tilde{r}^* & \tilde{t}^* \\
    \end{pmatrix} 
    \begin{pmatrix}
        \gamma_{00} & \gamma_{01} \\
        \gamma_{10} & \gamma_{11} \\
    \end{pmatrix}
	\end{equation}
\end{subequations}
is given up to a global phase.
The unitary part of the time evolution is described by the transmission and reflection coefficients $\tilde{t}(q,\tau)$ and $\tilde{r}(q,\tau)$ describing the transformation $\tilde{\mathcal{U}}_0$ on the main classes' subspace with
\begin{subequations}\label{eq:trans_refl_tilde}
\begin{align}
	\tilde{t} &= \ee^{\ii \tilde{\varphi}} \left[t(f) t(\beta f \varepsilon^2/8) - r(f) r^*(\beta f \varepsilon^2/8)\right],\\
	\tilde{r} &= \ee^{\ii \tilde{\varphi}} \left[t(f) r(\beta f \varepsilon^2/8) + r(f) t^*(\beta f \varepsilon^2/8)\right],
\end{align}
\end{subequations}
where $\tilde{\varphi}(\tau) = \Delta\omega \tau/(2\Omega_0)$ is the dynamical phase imprinted by the lasers, $\Delta\omega$ the laser detuning, $\Omega_0$ the Rabi frequency and the coefficients $t$ and $r$ are given by
\begin{subequations}
\begin{align}\label{eq:reflx_trans_coef}
	t(\Omega) = \cos \Omega\tau
	+ \ii\frac{v}{f}\sin \frac{\Omega\tau}{2}
    \quad\text{and}\quad
	r(\Omega) = -\ii\frac{\ee^{-\ii\theta}}{f}\sin \frac{\Omega\tau}{2}.
\end{align}
\end{subequations}
Here, $\theta$ denotes the phase of the Rabi frequency at the beginning of the pulse.
The non-unitary part of the time evolution is given by the coefficients $\gamma_{ij}(q, \tau)$, which can be brought into the form
\begin{align}
        \gamma_{01} &= \ee^{-\ii \theta} \varepsilon^2 \gamma_\mt{od}(-q,\tau),
        \quad
		\gamma_{00} = 1 + \varepsilon^2 \gamma_\mt{d}(-q,\tau),\\
		\gamma_{11} &= 1 + \varepsilon^2 \gamma_\mt{d}(q,\tau),
		\quad
		\gamma_{10} = \ee^{\ii \theta} \varepsilon^2 \gamma_\mt{od}(q,\tau),
\end{align}
where the \textit{diagonal} $\gamma_\mt{d}$ and \textit{off-diagonal} $\gamma_\mt{od}$ contributions are given by
\begin{subequations}
	\begin{align}
    \begin{split}
		\gamma_\mt{d}(q,\tau) = &
        - \frac{1}{16}-\frac{3 \ii v}{32 f^3}  \sin\! f \tau \\
        & + \frac{\ee^{-\ii\left(\frac{2}{\varepsilon}  + \frac{3v}{2} \right) \tau}}{16}\left( \cos \frac{f \tau}{2} + \ii \frac{v}{f} \sin \frac{f \tau}{2}\right)
    \end{split}
	\end{align}
    and
	\begin{align}
    \begin{split}
			\gamma_\mt{od}(q,\tau) = & \frac{3 v}{16f^2}\frac{1-v}{f-v }\sin^2 \frac{f \tau}{2}\\
            &+\frac{\ii  \ee^{ -\ii\left(\frac{2}{\varepsilon}  + \frac{3v}{2}\right)\tau }}{16 f} \sin \frac{f \tau}{2}+  \ii \frac{3 v^2}{32 f^3}\sin\! f \tau.
    \end{split}
	\end{align}
\end{subequations}
Note that $\gamma_\mt{d}$ and $\gamma_\mt{od}$ still contain trigonometric functions that depend on $\varepsilon$ to avoid secular terms as mentioned above.
In case of negligible coupling to the adjacent classes, the time evolution $G$ becomes unitary with $\gamma_{ij} = \delta_{ij}$ and the standard~\cite{Szigeti2012-Velocity_selectivity} coefficients $\tilde{t} = \ee^{\ii \tilde{\varphi}}t(f)$ and $\tilde{r} = \ee^{\ii \tilde{\varphi}}r(f)$.

Here, we also list the couplings to the adjacent momentum classes $m=-1,2$ starting from the main classes $n=0,1$. The corresponding field operators in the Heisenberg picture are given by 
\begin{equation}\label{aeq:adjacent_Psi_heisenberg}
    \hat{\Psi}_{m,\mt{H}}(q,\lambda)\ket{\psi_0} = \ee^{-\ii\omega_m\lambda} \sum_{n=0}^1 \gamma_{mn} \hat{\Psi}_{n}(q,0)\ket{\psi_0}
\end{equation}
up to a global phase, where $\omega_{m} = 2 + (-1)^m 3\varepsilon v / 2 + \varepsilon^2[2 - (-1)^m \varepsilon v]/8 + m\Delta\omega/\omega_k$ and $m=-1,2$. 
We assume that the initial state $\ket{\psi_0}$ only populates the main momentum classes $n=0,1$.
The coefficients $\gamma_{ij}(q, \tau)$ are given by
\begin{widetext}
\begin{subequations}
	\begin{align}
        \gamma_{20} &= \frac{\varepsilon \ee^{- 2 \ii \theta}}{16} \left( \varepsilon - \ee^{\ii \left(\frac{2}{\varepsilon} + \frac{3 v}{2} + \frac{\varepsilon}{4}\right) \tau} \left[ \varepsilon \cos \frac{f \tau}{2} -\ii \frac{4-3 \varepsilon v }{f} \sin \frac{f \tau}{2} \right]
        \right),\\
        \gamma_{-11} &= \frac{\varepsilon \ee^{2\ii \theta} }{16} \left( \varepsilon - \ee^{\ii\left(\frac{2}{\varepsilon}- \frac{3v}{2} + \frac{\varepsilon}{8}\right) \tau} \left[\varepsilon \cos\frac{(4f-\varepsilon)\tau}{8} - \ii \frac{4+3 v \varepsilon }{f}\sin\frac{(4f-\varepsilon)\tau}{8}  \right]\right),\\
        \gamma_{21}&=\frac{\varepsilon \ee^{-\ii\theta}}{16} \left\{\ee^{\ii\left(\frac{2}{\varepsilon} + \frac{3}{2}v + \frac{\varepsilon}{4}\right)\tau}
        \left[
            \frac{\ii f\varepsilon}{f(f+v)} \sin\frac{f\tau}{2}
            - \frac{4 - 3v\varepsilon }{f(f+v)} \cos\frac{f\tau}{2}
            - \ee^{-\ii\frac{f}{2}\tau}v\left(\varepsilon + \frac{4 - 3v\varepsilon }{f} \right)
        \right]
        + 4 - 2 v \varepsilon
        \right\},\\
       \gamma_{-10} &= \frac{\varepsilon \ee^{\ii\theta}}{16} \left\{\ee^{\ii\left(\frac{2}{\varepsilon} - \frac{3}{2}v + \frac{\varepsilon}{8}\right)\tau}
        \left[
            \frac{\ii f\varepsilon}{f(f+v)} \sin\frac{(4f-\varepsilon)\tau}{8}
            - \frac{4 + 3v\varepsilon }{f(f+v)} \cos\frac{(4f-\varepsilon)\tau}{8}
            + \ee^{\ii\frac{4f-\varepsilon}{8}\tau}
            v\left(
                    \varepsilon - \frac{4 + 3v\varepsilon}{f}
            \right)
        \right]
        + 4 + 2v\varepsilon\right\},
	\end{align}
\end{subequations}
\end{widetext}
where $\gamma_{20}$ and $\gamma_{-11}$, as well as $\gamma_{21}$ and $\gamma_{-10}$ are of the same structure and only differ in the dependence on the Doppler detuning and the trigonometric functions' frequencies.
\begin{figure*}[htb]
	\centering
	\includegraphics{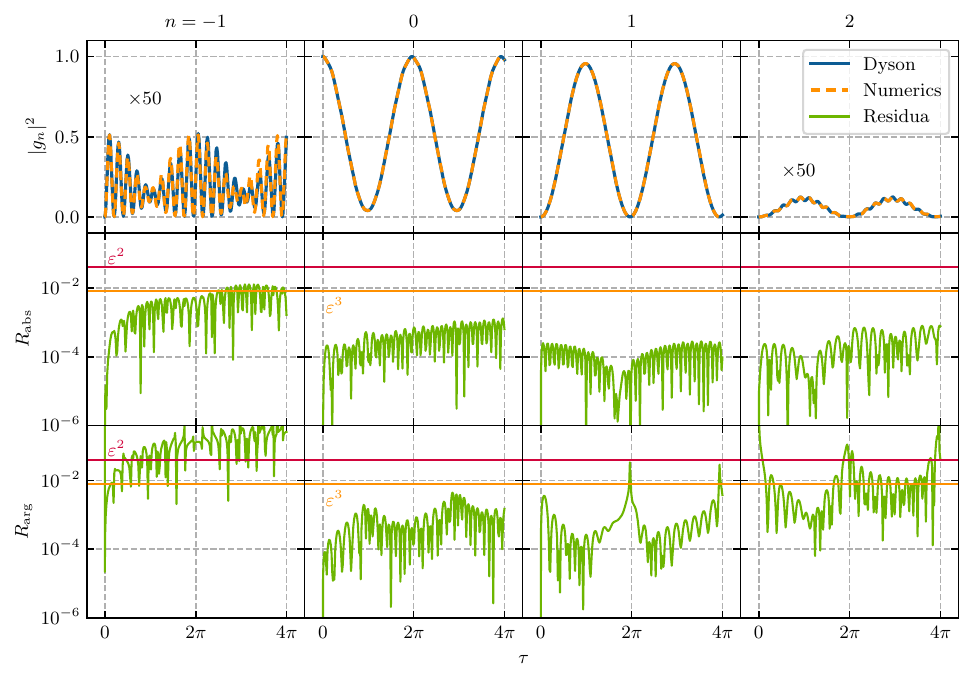}
	\caption{
        Comparison of the perturbative solution up to second order in $\varepsilon$ for first-order Bragg diffraction (solid blue) between the momentum classes $n=-1,0,1,2$ and the numerical model (dashed orange) detailed in Sec.~\ref{sec:num_sol}. 
        We consider the initial state $\ket{\psi_0}$, where an atom has the off-resonant momentum $p = 0.02 \hbar k$, \ie{} is in the zeroth momentum class in a momentum eigenstate, and the coupling strength is set to $\varepsilon = 0.2$.
    	For a first-order Bragg pulse with pulse area $\tau$, the population probability $|g_n|^2$ of the momentum classes (top row) exhibits Rabi oscillations between the main momentum classes, while there are atoms lost to the adjacent classes (factor $50$ for better visibility).
        Here, $g_n$ is given by Eqs.~\eqref{aeq:Psi_main_Heisenberg} and \eqref{aeq:adjacent_Psi_heisenberg} and obtained from $\hat{\Psi}_{n,\mt{H}}(q,\lambda)\ket{\psi_0} = g_n \hat{\Psi}_{0}(q,0)\ket{\psi_0}$.
        The residuals (green) of the perturbative calculation $g_n$ \wrt{} the corresponding numerical calculation $g_{n,\mt{num}}$ in modulus $R_\mt{abs} = ||g_n|-|g_{n,\mt{num}}||$ (center row) and phase $R_\mt{arg} = |\operatorname{arg}(g_n) - \operatorname{arg}(g_{n,\mt{num}})|$ (bottom row) show that the four-class perturbative calculation coincides with the numerical model in the main momentum classes $n=0,1$ to desired precision. 
        The peak in the phase residual on $n=1$ is merely a numerical artifact. 
        In contrast to the modulus, the phases of the adjacent classes are not accurately described, which is due to the truncation to four momentum classes in the perturbative treatment.
    }
	\label{fig:pops_and_residua_1stO_bragg}
\end{figure*}

Figure~\ref{fig:pops_and_residua_1stO_bragg} shows a comparison between the perturbative solution listed above and a numerical calculation solving first-order Bragg diffraction on six momentum classes $n=-2,-1,\ldots,3$ as outlined in Sec.~\ref{sec:num_sol}.
Here, the atom is initially in $n=0$, with a well-defined, sightly off-resonate momentum $p$, interrogated by a first-order Bragg pulse with variable pulse area $\tau$.
The top center panels show Rabi oscillations between the main momentum classes, while the top left and right panels confirm population transfer to the adjacent classes.
The population of the adjacent classes is suppressed, but oscillates at the same dominant frequency.
However, it is in addition fast oscillating, caused  by the strong detuning of the transitions between main and adjacent classes.
The residuals in modulus (center row) and phase (bottom row) confirm that the four-class perturbative description introduced above describes the coupling to the main momentum classes not only to the targeted precision, but even up to $\mathcal{O}(\varepsilon^4)$.
While the moduli of populations in the adjacent classes are accurately described, the perturbative solution fails to describe their phases. 
Because this effect is anticipated and arises from the truncation, such that including additional momentum classes would improve the description, the phases of the parasitic momentum classes are not relevant for our treatment, and our chosen truncation is therefore sufficient.
\vfill\eject

\section{MZI Phase Uncertainty for OAT States}
\label{app:phase_uncertainty_formulas}
In this appendix, we list properties of the OAT state $\ket{\chi_\alpha} = \ee^{-\ii \alpha \hat{S}_1}\hat{U}_\mt{oat}\ket{\tfrac{\pi}{2},0}_\mt{css}$, which is polarized in $x_1$ direction and denote the corresponding MZI phase uncertainty. 
In the OAT state's definition, we include a subsequent rotation around $x_1$ by an angle $\alpha$, which can be induced by a suitable laser pulse.
The MZI phase uncertainty, for interferometer and first-beam-splitter phase $\phi = 0 = \theta_0$, is given by Eq.~\eqref{eq:PS_4x4_Bragg_MZI} and depends on the angular momentum projections, covariances and variances.
The OAT state's projections are $\langle \hat{S}_{1} \rangle = \cos^{N-1}\!\chi N/2$ and $\langle \hat{S}_{2} \rangle = 0 = \langle \hat{S}_{3} \rangle$, while the variances are given by~\cite{Kitagawa1993-SSS}
\begin{subequations}\label{aeq:equator_OAT_moments}
\begin{align}
	\Delta S_{1}^2
	= & \frac{N}{4}\left[N\left(1-\cos^{2N-2}\chi\right)-\frac{N-1}{2}A\right],
	\\
	\Delta S_{2/3}^2
	=& \frac{N}{4} \left\{
	1 + \frac{N - 1}{4} 
	\left[A \pm \sqrt{A^2 + B^2 \cos(2\varphi)} \right]
	\right\},
\end{align}
and the covariances by
\begin{align}
	\operatorname{Cov}[\hat{S}_1, \hat{S}_2] &= 0 = \operatorname{Cov}[\hat{S}_1, \hat{S}_3],\\
\begin{split}
	\operatorname{Cov}[\hat{S}_2, \hat{S}_3] &= 
	\frac{N(N - 1)}{4}\Big\{
	\cos [2 (\varphi-\alpha_0)]
	\sin \chi
	\cos^{N-2} \chi\\
     &\quad+
    \frac{\sqrt{A^2 + B^2}}{4} \sin [2(\varphi-\alpha_0)]\cos (2 \alpha_0) \Big\}.
\end{split}
\end{align}
\end{subequations}
Here, we have introduced the abbreviations $A = 1 - \cos^{N - 2} 2\chi$ and $B = 4 \sin\chi \cos^{N - 2}\chi$, which give rise to the initial inclination $\alpha_0 = \arctan( B/A )/2$ of the OAT ellipse \wrt{} the equator and strongly depend on twisting strength $\chi$ and atom number $N$.
We furthermore introduce the final inclination $\varphi = \alpha + \alpha_0$ of the OAT state after the rotation about $\alpha$ around $x_1$.
Equation~\eqref{aeq:equator_OAT_moments} shows how the (co)variances are highly sensitive to changes in this inclination.
Minimal variance along $x_3$ is achieved for $\varphi = 0$, \ie{} $\alpha = -\alpha_0$.
Using Eq.~\eqref{eq:PS_4x4_Bragg_MZI} and the OAT state's properties, the MZI phase uncertainty for $\phi=0=\theta_0$ is given by
\begin{align}\label{aeq:ps_alpha_equator_OAT}
\begin{split}
		\Delta\phi^2&|_{\phi,\theta_0=0} = \biggl\{\left(\Re\mathscr{A}_{10}\right)^2\Delta S_1^2 
		+ \left(\Im\mathscr{A}_{10}\right)^2\Delta S_2^2 \\
		&+ \mathscr{A}_{0}^2\Delta S_3^2 
		+ 2\mathscr{A}_{0} \Im\mathscr{A}_{10}\operatorname{Cov}\bigl[\hat{S}_2, \hat{S}_3\bigr]\\ 
        &\left. + \frac{N}{4}V_\mt{M} + \frac{\mathscr{R}_{10}}{2} \langle\hat{S}_{1}\rangle \biggr\}
        \middle/ 
        \left(\Re\mathscr{A}'_{10} \langle \hat{S}_1 \rangle 
        - \mathscr{A}_{0}'\frac{N}{2} \right)^{2} \right.,
\end{split}
\end{align}
where the script quantities and $V_\mt{M}$ are dependent on the coupling strength $\varepsilon$ and the momentum-mode function $\varphi_0$. They are listed in Sec.~\ref{sec:PS_including_VS_and_PD} and Table~\ref{tab:scr_quantities}.

\bibliography{literature.bib}

\end{document}